# Composition Heterogeneity Induced Crystallization in Double Crystalline Binary Polymer Blends

Ashok Kumar Dasmahapatra[*]

Department of Chemical Engineering, Indian Institute of Technology Guwahati,

Guwahati - 781039, Assam, India

---

[*] Corresponding author: Phone: +91-361-2582273, Fax: +91-361-2582291, Electronic mail: akdm@iitg.ernet.in

**Abstract**

Polymer blends offer an exciting material for various potential applications due to their tunable properties by varying constituting components and their relative composition. Our simulation results unravel an intrinsic relationship between the phase behavior and crystallization characteristics with the relative composition of A- and B-polymer in the system. We report simulation results for non-isothermal and isothermal crystallization with weak and strong segregation strength to elucidate the composition dependent crystallization behavior. With increasing composition of low melting B-polymer, macrophase separation and crystallization temperature changes non-monotonically, which is attributed to the change in diffusivity of both the polymers with increasing composition of B-polymer. In weak segregation strength, however, at high enough composition of B-polymer, A-polymer yields relatively thicker crystals, which is attributed to the dilution effect exhibited by B-polymer. When B-polymer composition is high enough, it acts like a solvent while A-polymer crystallizes. Under this situation, A-polymer segments become more mobile and less facile to crystallize. As a result, A-polymer crystallizes at a relative low temperature with the formation of thicker crystals. At strong segregation strength, dilution effect is accompanied with the strong A-B repulsive interaction, which is reflected in non-monotonic trend of mean square radius of gyration with increasing composition of B-polymer. Isothermal crystallization reveals a strong relationship between composition and crystallization behavior. Two-step (viz., sequential crystallization) yields better crystals than one-step (viz., coincident crystallization) for both the polymers.

## 1. Introduction

Polymer blends (a mixture having two or more polymers)[1,2], possess superior properties over pure polymer for a wide range of modern applications such as, nanoelectronics[3], polymer-based light emitting diodes[4] (LED) and medical appliances.[5] In most of the cases, constituting components are chemically dissimilar and hence phase separate via macrophase separation, which may be overcome by application of a suitable compatabilizer to bring miscibility. The extent of phase separation dictates the overall morphological development, influencing various properties of the resultant blend. A miscible binary blend shows a single glass transition temperature[6-14] ($T_g$), whereas an immiscible blend shows two $T_g$s, corresponding to the respective components.[15,16]

Blends with one crystallizable component have been studied extensively to understand miscibility pattern, phase behavior and crystallization characteristics. Crystallization behavior of binary blends with two crystallizable components are challenging due to the complexity arises from the interplay between macrophase separation driven by mutual immiscibility, and crystallization. Usually, the high melting component (HMC) crystallizes first and form crystalline domain, creates confinement for the crystallization of low melting component (LMC), which crystallizes in the confined space created during the crystallization of HMC. In most of the cases the crystal domains are separated from each other, but in some cases, interpenetrating[17-21] and mixed crystals are observed.[22]

Mutual miscibility plays a pivotal role in determining phase behavior and morphological development in binary blends. Close proximity in the melting points of the components usually facilitates forming a miscible blend, exhibiting a single $T_g$[6-14] and interpenetrating spherulites.[17-19] However, this is not always true. For example, Poly(ethylene oxide) (PEO) and Poly(ε-caprolactone) (PCL) have melting points close to each other ($T_m \sim 70$ °C), even though they produce immiscible blends.[23] Molecular weight of the constituting components also influences miscibility. For example, in the blend of Poly(3-hydroxybutyrate) (PHB) ($T_m$: 176 °C) and Poly(L-lactic acid) (PLLA)[24], high molecular weight (MW) PLLA ($T_m$: 176 °C) produces an immiscible blend whereas, low MW ($T_m$: 123 °C) produces a miscible blend. Therefore, the extent of interaction between the components plays a major role in deciding the mutual miscibility. Immiscible blends usually show two distinct glass transition temperatures ($T_g$s) for individual components; however the values of $T_g$ are influenced by the constituting components.

Apart from the melting point difference between two components and mutual immiscibility, the relative composition of the constituting components plays a critical role in determining phase behavior, crystallization mechanism and morphological development. Increasing composition heterogeneity (viz., one component is higher in proportion than the other), leads to changing in glass transition, melting and crystallization temperature affecting crystal morphology. For example, in the blends of poly(ethylene suberate) (PESub) and (PEO),[25] where the melting and glass transition temperature of the components are close to each other (PESub: $T_m$ ~ 62 °C, $T_g$ ~ - 48 °C and PEO: $T_m$ ~ 64 °C, $T_g$ ~ -51 °C), the crystallization behavior is largely governed by the relative composition of the components. Major component crystallizes first and minor component is either expelled from the crystalline domain or included in the inter-lamellar space – thus, a fractional crystallization of minor component is observed.

A variety of phenomenon has been observed in the crystallization of asymmetric binary polymer blends. For example, in the blends of Poly(3-hydroxybutyrate) (PHB) and PEO,[26] increasing % of PEO leads to an increase in melting and glass transition temperature of PHB, whereas melting temperature of PEO decreases and a fractionated crystallization of PEO has been observed.[27] Isothermal crystallization reveals that crystallization rate of PHB decreases in the presence of PEO.[6] Blends of PHB and Poly(L-Lactic Acid) (PLLA), shows composition dependent[24] and PHB molecular weight (MW) dependent[8] miscibility behavior. Melting point of PLLA increases in the presence of PHB;[28] lamellar thickness increases in the presence of ataPHB[8], which is incorporated within the inter-lamellar region of PLLA spherulites. Crystallization mechanism also changes depending on the composition: for symmetric blend it follows a simultaneous crystallization whereas, in asymmetric blend, it follows a sequential crystallization pathway.[24] Usually, simultaneous crystallization behavior is shown by the blends where $T_m$s of the components are closed enough.[29] However, in the blend of PHB and PLLA, the 50:50 miscible blend follows a simultaneous crystallization mechanism, although the rate of crystallization of PLLA is faster than PHB, and the final crystallinity of PLLA is less due to the simultaneous crystallization.[24] In the blend of Polypropylene (PP) and PLA,[5] melting temperature of PP decreases by ~ 5 °C on increasing PLA composition from 40 – 60%. On the other hand, in the blend of PLLA and Poly(ethylene succinate) (PES),[30] being an immiscible blend, $T_g$ and $T_m$ appears to be independent of composition. However, the rate of crystallization of PLLA is increased with increasing PES content, which is attributed to the fact that the interface acts as a nucleating agent for the crystallization of PLLA. In contrary, the rate of crystallization of PES is decreased with increasing PLLA content due to the fact that the crystallization of PLLA happens under confinement created during the crystallization of PLLA. In the blend of poly(butylene succinate) (PBS) and PEO[31], with increasing % of PEO, crystallization of PBS is inhibited with a melting point depression. On the other hand, crystallization rate of PHBV decreases with increasing PBS content.[32] In the blend of PBS and PCL, The crystallization of PBS is almost unaffected by PCL.[33] However, crystallization of PCL exhibits two opposite trends with increasing PBS – first increased and then decreased with increasing the content of PBS. This unexpected trend in crystallization behavior is attributed to the fact that initially formed crystalline domains of PBS act as nucleating agent, enhancing the crystallization of PCL, but subsequently, due to the confinement effect the crystallization of PCL is decreased. In the blend of PEO and PCL,[23] which form an immiscible and biphasic melt, crystallization rate of PEO decreases with increasing % of PCL, however, crystallization of PCL is almost remain unaffected due to change in relative composition. In the blend of PEO and PES,[22] PES being the HMC ($T_m \sim 101$ °C) crystallizes first followed by the crystallization of PEO (LMC, $T_m \sim 59$ °C). When PEO is added $\leq 20\%$, the crystallization kinetics of PES is enhanced, which is attributed to the dilution effect due to which the chain mobility increases. However, a reduction in crystallization driving force is observed when PEO is increased beyond 20%. Kinetic analysis reveals a path-dependent (viz., one-step and two-step cooling) crystallization behavior. In one-step cooling, usually the crystallization mechanism follows a simultaneous crystallization, where both the components crystallize simultaneously but with a different rate of crystallization.[24] However, if the rate of crystallization of one component (usually, HMC) is fast enough to fill the entire space before the other component (LMC) starts to crystallize, then the mechanism changes to sequential crystallization, typically observed in two-step cooling.[9]

Recently, we have demonstrated that increase in the segregation strength leads to the formation of smaller and thinner crystals with less crystallinity in a symmetric binary polymer blend.[34] This paper reports simulation results of crystallization of binary polymer blend of two crystallizable components to explore the effect of relative composition of the constituting components on crystallization. Our results show a non-monotonic trend in crystallinity and lamellar thickness with increasing composition of B-component. Isothermal experiments reveal a strong dependency of the composition heterogeneity on transition pathways.

We organize our paper as follows. In the section 2, we report our model and simulation technique. Following this, we discuss our key results in the section 3 and summarize our results in the section 4.

## 2. Model and Simulation Technique

We use a coarse-grained lattice model of A- and B-polymers each with 64 coarse-grained units. In a cubic lattice of size 32×32×32, we put total 480 chains consisting NCA number of A- and NCB number of B-polymers, along the lattice grids. Thus, the occupation density becomes 0.9375, representing a bulk polymer system. To implement variation in composition in the blend, we vary the composition of B-polymer, $x_B$ ranging from 0.125 to 0.875 with an interval of 0.125. Accordingly, numbers of A- and B-polymer chains are decided. For example, if $x_B$ = 0.25, then NCB = 480×0.25 = 120, remaining chains (viz., 480×0.75 = 360) are of A-polymer. In the crystallization process, crystallization is facilitated by crystallization driving force, whereas the chemical dissimilarity between A- and B-units promotes immiscibility between them and leads to a phase separated state. We model crystallization driving force as an attractive interaction between neighboring parallel bonds and collinear bonds within A- or B- type units, represented by $U_p$ and $U_c$ respectively. The interaction between A- and B-type units is modeled as a repulsive interaction represented by $U_{AB}$. The change in energy per Monte Carlo move is then:

$$\Delta E = -\left(\Delta N_p U_p + \Delta N_c U_c\right)_A - \left(\Delta N_p U_p + \Delta N_c U_c\right)_B + \Delta N_{AB} U_{AB}$$

Where, $\Delta N_A$ and $\Delta N_B$ represents the net change in the number of parallel and collinear bond respectively for the A and B polymer, and $\Delta N_{AB}$ represents the change in the number of contacts between A and B units.

We model A-polymer as the high melting component (HMC), thus it will crystallize before B-polymer during crystallization from a high temperature homogeneous melt, followed by low-melting B-polymer, which is less facile to crystallize. Therefore, the crystallization driving force for B-polymer would be relatively less compared A-polymer. To implement this, we take $U_{pB} = \lambda_m U_{pA}$ and $U_{cB} = \lambda_m U_{cA}$ for the parallel and collinear bond interaction energy respectively, and we set $\lambda_m = 0.75$ (<1) to represent that B-polymer has less driving force for crystallization compared to A-polymer. Further, we assume that $U_p = U_c$, for the coarse grained interactions (in lattice model) used in our simulation. To implement the mutual immiscibility (in the form of repulsive interaction between A- and B-units) we use $\lambda$ as the measure of segregation strength (viz., A-B demixing energy), which is equivalent to Flory's $\chi$ parameter. Accordingly, the interaction between A- and B-polymer, $U_{AB}$ is represented as $\lambda U_p$. In this work we simulate systems with two segregation strengths: we set $\lambda = 1$ and 6 for weakly and strongly segregated systems, respectively. In terms of Flory's $\chi$ parameter, segregation strength is calculated as $\chi N$, which may be correlated to $(q-2) \times U_{AB} \times N$ in our system,[35-37] where $q$ is coordination number and $N$ is the degree of polymerization. All the energies are normalized by $k_B T$, where, $k_B$ is the Boltzmann constant and $T$ is temperature in Kelvin; thus, $U_p \sim 1/T$. Now the change in energy per MC move is modified as follows:

$$\Delta E = \left[ -\left( \Delta N_p + \Delta N_c \right)_A - \lambda_m \left( \Delta N_p + \Delta N_c \right)_B + \lambda \Delta N_{AB} \right] U_p$$

We apply a set of microrelaxation moves to simulate polymer chains in the lattice. We employ single site bond fluctuation algorithm along with periodic boundary conditions to move chain molecules along the lattice grids. The coordination number of our cubic lattice is 26 (6 along the lattice axis, 12 along the face diagonals and 8 along the body diagonal). Thus, bond length can be 1 (along the axis), $\sqrt{2}$ (along the face diagonal) or $\sqrt{3}$ (along the body diagonal) lattice units. We start our simulation by selecting a vacant site randomly from the available vacant sites and searching for a nearest site occupied by either A- or B-type units. Once an occupied site is found, an appropriate microrelaxation move is selected according to their position along the chain − for terminal units, end bond rotation and slithering diffusion is selected with equal probability; for non-terminal units, a single site bond fluctuation is implemented.[34,38] During the movement we have strictly implemented the excluded volume effect − one lattice site is occupied by one unit (either A- or B-type), and no bond crosses each other. Once this self-avoiding criterion is satisfied, we calculate the change in energy associated with the movement.

We employ the Metropolis sampling scheme to sample new conformations. The probability of an MC move is given by $\exp(-\Delta E)$. We accept new conformation if $\exp(-\Delta E) \geq r$, where $r$ is the random number in the range $(0, 1)$, generated by using the random number generator MT19937.[39] We simulate the crystallization of binary blend by varying $U_p$ from 0 (viz., at infinite temperature, athermal state) to 0.6 with a step size 0.02, to represent a step-cooling from a high temperature melt. To equilibrate the system we calculate mean square radius of gyration, $\langle R_g^2 \rangle$, as a function of Monte Carlo Steps (MCS). Variation of $\langle R_g^2 \rangle$ with MCS does not show appreciable change beyond 5000 MCS and it is considered as the equilibration time. We calculate thermodynamics and structural parameters averaged over subsequent 5000 MCS.

To monitor transition from a disordered melt to an ordered crystalline phase, we calculate specific heat ($C_v$) and fractional crystallinity, $X_c$ of A-polymer, B-polymer and overall as a function of $U_p$. Specific heat is calculated as equilibrium specific heat from the total energy fluctuations (for all the A- and B-type units in the simulation box), similar to that of Dasmahapatra et al.[34,38] We define crystallinity as the ratio of numbers of crystalline bonds to the total numbers of bonds present in the system. A bond is considered to be crystalline if it is surrounded by more than 5 nearest non-bonded parallel bonds.[34,36,37] To locate the macrophase separation point, we calculate $C_v$ for A-B pair ($C_{v\_AB}$) based on the de-mixing energy between A and B-polymer.[34] During macrophase separation, individual polymer forms their respective domains, and the resulting transition may be considered as a disordered to ordered transition, which would exhibit a peak in $C_{v\_AB}$ at the transition point. We also calculate mean square radius of gyration, $\langle R_g^2 \rangle$, average crystallite size, $\langle S \rangle$ and lamellar thickness, $\langle l \rangle$ as a function of $U_p$. A crystallite size $S$ is defined as a microscopic aggregate having $S$ numbers of crystalline bonds (A- or B-type) in the same direction. We express lamellar thickness as the average number of monomer units (A- or B-type) present in a given crystallite, and averaged over all the crystallites present in the system.

### 3. Results and Discussions

We begin with describing the details of non-isothermal crystallization of A/B binary polymer blends containing total 480 polymer chains exploring the effect of composition (viz., $x_B = 0.125, 0.25, 0.375, 0.5, 0.625, 0.75$ and $0.875$) on crystallization and morphological development. We discuss the composition effect at weak (viz., $\lambda = 1$) and strong (viz., $\lambda = 6$) segregation strength. Subsequently, we discuss isothermal crystallization (one- and two-step) with the detail analysis on development of crystal morphology with both the segregation strengths.

### 3.1. Non-isothermal Crystallization
### 3.1.1. Monitoring Phase Transition

For each composition, we prepare a homogeneous melt of A/B binary blend by equilibrating the sample system at $U_p = 0$ (T = ∞, athermal state). Following this, we gradually cool the sample system to $U_p = 0.6$ with a step size 0.02. The step-wise cooling experiment corresponds to a non-isothermal crystallization process. Homogeneous melt at $U_p = 0$ is well characterized by isotropic distribution of A- and B-segments throughout the system. Figure 1 represents snapshots for $x_B = 0.25$ and 0.75 showing an evenly dispersed melt system of binary blend (blue and magenta lines represent A- and B-polymers respectively). Snapshots for other compositions are available in supplementary information, Figure S1.[40] As the system is cooled by increasing $U_p$ (viz., decreasing temperature) polymer chain segments start arranging in parallel orientation. Increasing $U_p$ leads to the transition from a disordered melt to an ordered crystalline state. We monitor the crystallization of binary melt by following the change in equilibrium specific heat,[34,41] calculated from the energy fluctuation as a function of $U_p$. The value of $U_p$ at which $C_v$ shows a peak is regarded as the transition point from melt to crystal state in tune with the experimental observation.[42] Figure 2a shows the variation in $C_v$ with $U_p$ for a series of composition: $x_B = 0.125$ to 0.875 for $\lambda = 1$ (weak segregation). Since A- and B-polymers have different melting points, we see two transition points for A- and B-polymer respectively. First peak corresponds to the crystallization of A-polymer followed by the crystallization of B-polymer as shown by the second peak. Thus, the crystallization of the blend follows a sequential crystallization mechanism. These results are in close agreement with experimental observations on the blends of PBS/PEO[31], PHB/PEO[26] and PHB/PBS[43], wherein the individual component shows the respective transition temperatures during melting and crystallization. Similar trend of $C_v$ vs. $U_p$ has also been observed in case of $\lambda = 6$ (strong segregation), shown in Figure 2b. Insets of Figure 2a and 2b display the change in transition point, in terms of $U_p$ ($U_p^*$), of A- and B-polymer with change in composition, $x_B$ for weak ($\lambda = 1$) and strong ($\lambda = 6$) segregation strength, respectively. At high segregation (viz., $\lambda = 6$), the transition point follows a non-monotonic trend with composition, xB, which is attributed to the strong repulsive interaction between components. At weak segregation (viz., $\lambda = 1$), transition points (in terms of $U_p$) of A-polymer remain almost unaffected with increasing composition, except for $x_B = 0.875$; however, there is no effect on B-polymer over the entire composition range investigated. Crystallization happens from a macrophase separated melt and hence individual component crystallizes within their own domains and almost unaffected by the other. However, for xB = 0.875, the transition point of A-polymer shifts towards higher value of $U_p$ (viz, lower temperature) due to the dilution effect shown by B-polymer. At higher $x_B$ (viz., $x_B = 0.875$), B-polymer acts like a "solvent" during the crystallization of A-polymer. We observe that when A-polymer crystallizes at $U_p \sim 0.28$, the mean square displacement of center of mass ($d_{cm}^2$) of A-polymer increases at higher $x_B$ (Figure 3a, weak segregation). Higher mobility of chain segments makes them less facile to crystallize at that temperature with the available thermodynamic driving force. On lowering the temperature further (viz. increasing $U_p$), thermodynamic driving force increases and overcomes the inertia caused due to the enhanced mobility. As a result, A-polymer crystallizes at a relatively low temperature (viz., depression of crystallization temperature). On the other hand, crystallization temperature of B-polymer is much less sensitive to the change in composition, since B-polymer crystallizes after A-polymer crystallization, within the macropahse separated domains. Therefore, there is practically no influence from A-polymer and/or composition of A/B to influence the crystallization. Hence, B-polymer crystallization temperature remains almost same (Figure 2a). This observation is in the same line with the experimental results on blends of PP and PLA,[5] which exhibit decrease in melting temperature and crystallinity of PP with increasing PLLA content from 40 – 60%. Similarly, in PEO/PES[22] blend, ~20% of PEO enhances the crystallization of PES due to the enhanced chain mobility; in the blend of Poly(butylene adipate-co-butylene succinate) (PBAS)/PEO,[29] both the component shows melting point depression. In the blend of PBS/Poly(vinylidene fluoride) (PVDF),[44] growth rate of PVDF decreased with increasing PBS content while PBS shows a melting point depression. Similar phenomena is also observed in the blend of PVDF and Poly(butylene succinate-co-butylene adipate) (PBSA).[45] In strongly segregated system (viz., $\lambda = 6$), a similar trend in center of mass diffusion has been observed (Figure 3b), which explains the trend in $U_p^*$ with increasing $x_B$ (Figure 2 insets).

### 3.1.2. Locating macrophase separation point

Immiscibility between the components leads to a macrophase separated structure in the melt. Therefore, when we cool our sample system from $U_p = 0$, they form a macrophase separated melt state, before crystallization. We estimate $C_{v\_AB}$ based on the interaction between A- and B-units to locate the macrophase separation point for all the compositions investigated. The value of $U_p$ at which $C_{v\_AB}$ shows a peak is considered as the macrophase separation point ($U_p^\#$). With increasing $x_B$, $U_p^\#$ shows a non-monotonic trend in both $\lambda = 1$ (Figure 4a) and 6 (Figure 4b). It appears from the trend that the asymmetry in composition retards the macrophase separation to a lower temperature (viz., higher $U_p$ value). The values of $U_p^\#$ at $\lambda = 6$ are ~ one order of magnitude smaller than that of at $\lambda = 1$. In both the cases, the system with composition asymmetry (viz., one component is more in amount than the other), the macrophase separation happens at a relatively low temperature (viz., higher value of $U_p$). These results can easily be correlated with the chain mobility, measured in terms of mean square displacement of center of mass, as presented in Figure 3. As the composition asymmetry increases, the mobility of minor component increases, and as a result, the macrophase separation is retarded and happens at a relatively lower temperature (viz., higher $U_p$ values). Once the macrophase separation occurs, the individual components (A and B) crystallize within the respective domains and with suppressed crystallization as seen in $C_v$ trend (Figure 2). Figure 5 shows snapshots of macrophase separated melt for Up

= 0.1 for $\lambda$ = 1 and 6 for $x_B$ = 0.25 and 0.75 respectively. Snapshots of the rest of the compositions are available in Figure S2, Supplementary information.[40]

We have calculated mean square radius of gyration, $\left\langle R_g^2 \right\rangle$ as a function of $U_p$ for both the segregation strengths and all the compositions, to get an insight in the demixing behavior of binary blends. Figure 6 shows that as the temperature decreases (viz., $U_p$ increases), we observe a sudden decrease in the value of $\left\langle R_g^2 \right\rangle$, which signifies the onset of macrophase separation, decreases further till $U_p$ = 0.28, where A-polymer starts to crystallize. Beyond $U_p$ = 0.28, $\left\langle R_g^2 \right\rangle$ possesses almost a constant value after a marginal increase, where B-polymer crystallizes in the confined space created during the crystallization of A-polymer. The trend is similar for both $\lambda$ = 1 (Figure 6a) and 6 (Figure 6b). The insets of Figure 6 displays the value of $\left\langle R_g^2 \right\rangle$ at $U_p$ = 0.28 (where A-polymer crystallizes) and $U_p$ = 0.6 (at the end of crystallization of both A- and B-polymer) as a function of $x_B$. Two opposite scenarios have been observed for $\lambda$ = 1 and 6. At $\lambda$ = 1, as the composition heterogeneity decreases (viz., $x_A \sim x_B$), the value of $\left\langle R_g^2 \right\rangle$ shows a decreasing trend, whereas, when either $x_A > x_B$ or $x_B > x_A$, the values are relatively large. This is attributed to the dilution effect in the presence of higher degree of composition heterogeneity. As discussed before, composition asymmetry retards macrophase separation; the formation of individual domains of minor component is less facile, which causes the enhancement of $\left\langle R_g^2 \right\rangle$. However, as $x_A \sim x_B$, the formation of individual domains and macrophase separation becomes more probable. As a result, the value of $\left\langle R_g^2 \right\rangle$ decreases. On the other hand, at $\lambda$ = 6, we see an opposite scenario. When $x_A \sim x_B$, the value of $\left\langle R_g^2 \right\rangle$ shows an increasing trend. This unexpected trend is attributed to the the strong repulsive interaction between the components. As the composition of both the components is relatively large, formation of large size individual domains is less facile in the presence of strong segregation strength. Large numbers of smaller size domains, which are interconnected (viz., a chain is a part of several domains) are formed. As a result, the chains are relatively stretched and exhibit a higher value of $\left\langle R_g^2 \right\rangle$. The trend at $U_p$ = 0.28 is retained at $U_p$ = 0.6 too, which signifies that the morphology set during the macropahse separation and subsequent A-polymer crystallization, is retained (viz., unperturbed) during crystallization of B-polymer, which happens in the confined space created during the crystallization of A-polymer (see Figure S3, Supplementary information,[40] for the snapshots at $U_p = 0.28$, where A-polymer is almost crystallized, B-polymer is still in molten state).

### 3.1.3. Development of Crystallinity

We study the development of crystallinity during non-isothermal crystallization by calculating crystallinity of A-polymer ($X_A$), B-polymer ($X_B$) and overall ($X$) as a function of $U_p$, for a series of $x_B$. Overall crystallinity ($X$) is calculated as an weighted average of summation of A- and B-components: $X = X_A x_A + X_B x_B$. Figure 7a shows an abrupt increase in $X$ ($\lambda = 1$) at $U_p \sim 0.28$ for all the values of $x_B$, including $x_B = 0$ (pure A-homopolymer) and reaches at a saturation crystallinity at c $\sim 0.5$. The saturation crystallinity of A-polymer, B-polymer and overall are presented in Figure 7b as a function of $x_B$. With increasing $x_B$, $X_B$ increases while $X_A$ decreases: as a result, overall crystallinity slightly increases at higher value of $X_B$, and the change in overall crystallinity $\sim 73 - 77\%$ (crystallinity of pure A-homopolymer $\sim 79\%$). In weak segregation limit, crystallization behavior and crystallinity is not drastically affected by either component. Degree of cooling is primarily responsible for the initial development of crystals. Saturation crystallinity of A- and B-polymers happens due to the change in relative composition. The values of saturation crystallinity of A- and B-polymer at $\lambda = 1$ is given in Table 1. Crystallinity of B-polymer increases marginally due to higher composition. But at low $x_B$, crystallinity of B-polymer is comparatively less compared to A-polymer, due to the fact that B-polymer crystallizes in the confined space created during macrophase separation and subsequent crystallization of A-polymer. At weak segregation, chain mobility of B-polymer shows a higher value (see Figure 3a) when A-polymer crystallizes ($U_p \sim 0.28$) across the compositions, since it is still in the molten state. As a result, crystallization is not hindered too much in weakly segregated system. Therefore, the saturation crystallinity of both the polymers lies within a narrow range with respect to the change in composition. However, in strongly segregated system (viz., $\lambda = 6$), we see a significant change in crystallinity as we increase $x_B$ (Figure 8). Saturation values of $X_A$, $X_B$ and $X$ show a non-monotonic behavior as a function of $x_B$. The values of saturation crystallinity of A- and B-polymer at $\lambda = 6$ show significant change compared to that of $\lambda = 1$ (see Table 1). This change in crystallinity is attributed to the effect of the segregation strength. At strong segregation strength, the interface becomes rigid with the formation of large numbers of smaller size domains, producing less crystalline material compared to $\lambda = 1$. We also observed that at strong segregation strength (viz., $\lambda = 6$), the chain mobility of both the polymers are significantly reduced (see Figure 3b), affecting development of crystalline domains. With increase in $x_B$, $X_B$ first decreases, and then increases. As the composition of each component is approaching equal to each other (viz., $x_A \sim x_B$), they compete with each other in the formation of individual domains. As a result, large numbers of smaller size domains are formed. These domains are interconnected and some of the chains are stretched further, which is evident from the increased value of mean square radius of gyration (Figure 6). The composition dependent crystallinity development has been observed in several blends such as in the blend of PLA and Poly(oxymethylene) (POM),[16] with closely spaced melting points, the easily crystallizable POM crystallizes first and restricts the crystallization of PLA when POM is more than 50% . Similarly, in the blend of PBS and PEO.[9,31] PBS crystallized first and the crystallization of PEO is restricted in the presence of already crystallized PBS. Figure 9 displays the snapshots of the crystalline structure at $U_p = 0.6$ for $x_B = 0.25$ and 0.75, $\lambda = 1$ and 6, respectively. Snapshots of rest of the compositions are available in Figure S4, Supplementary information.[40]

### 3.1.4. Structural analysis

To get an insight into the structural evolution during crystallization, we estimate average crystallite size and lamellar thickness for both the components as a function of $U_p$ at $\lambda = 1$ and 6, for all the compositions. Crystallite size shows a wider distribution compared to that of lamellar thickness, and the magnitude of lamellar thickness is much smaller in comparison with crystallite size − it indicates the formation of two dimensional crystals during crystallization. The variation of $\langle S \rangle$ with $U_p$ for A- and B-polymer are available in Figure S5, Supplementary information.[40] Figure 10a and b displays the variation of $\langle S \rangle$ with composition, $x_B$ at $U_p = 0.6$ for A- and B-polymer at $\lambda = 1$ and 6, respectively. In both the segregation strengths, the crystallite size of A- and B-polymers shows a non-monotonic trend. At $\lambda = 1$ (viz., weak segregation strength), the variation of $\langle S \rangle$ is relatively small compared to that of at $\lambda = 6$. At $\lambda = 6$, with increasing composition asymmetry, the value of $\langle S \rangle$ for both the components increases, which can be correlated with the chain mobility (see Figure 3) and dilution effect. At $\lambda = 6$, due to strong repulsive interaction between the components, the interface becomes more rigid compared to $\lambda = 1$. As a result, the average size at $\lambda = 6$ is much smaller compared to that of $\lambda = 1$. The formation of larger size crystallites is more probable at $\lambda = 1$ compared to that of at $\lambda = 6$. The average crystallite size at weak segregation strength, thus shows an irregular trend compared to that of at $\lambda = 6$.

We calculate average lamellar thickness as a function of $U_p$ for all the compositions investigated. The overall trend looks similar to that of crystallinity (Figure S6, Supplementary information).[40] As expected the saturation value of $\langle l \rangle$ is higher for $\lambda = 1$ compared to that of $\lambda = 6$ for a given composition. Figure 11a and b represent the variation of saturation value of $\langle l \rangle$ (viz., value at $U_p = 0.6$) as a function of $x_B$ for $\lambda = 1$ and 6, respectively. The saturation value of $\langle l \rangle$ for $\lambda = 1$ shows that $\langle l \rangle$ for B-polymer, $\langle l_B \rangle$ exhibit an overall decreasing trend with increasing $x_B$. For A-polymer, $\langle l_A \rangle$ increases at higher value of $x_B$. This non-intuitive trend in lamellar thickness is attributed to the dilution effect shown by B-polymer. Crystallization mechanism follows different pathway as the relative composition of the blend changes. When $x_B$ is low, the crystallization of A-polymer follows a typical "melt crystallization' mechanism, where it is influenced by intra- and inter-chain entanglement. However, at higher value of $x_B$, during the crystallization of A-polymer, B-polymer is still in the molten state. A-polymer crystallizes in a molten matrix of B-polymer, which acts like a "solvent" for A-polymer, reduces the hindrance and facilitates the formation of extended chain crystals. In the melt crystallization mechanism, folded chain crystals are preferred. Due to the formation of extended crystals, crystal thickening takes place with the formation of thicker crystals. When $x_B << x_A$, $\langle l_B \rangle > \langle l_A \rangle$. This can be explained as follows: when B-polymer starts crystallizing, A-polymer is already crystallized, therefore, B-polymer is not experiencing any hindrance from A-polymer, facilitating crystal thickening of B-polymer. As $x_B$ increases, due to the entanglement effect, the segmental mobility is reduced and as a result, relatively thinner crystals formed. At very high $x_B$, due to the presence of large composition thicker crystals are formed with higher crystallinity. A similar non-monotonic trend has also been observed in case of strong segregation strength ($\lambda = 6$) (Figure 11b). Thus, it can be inferred from the above analysis that the dilution effect dominate in the presence of higher degree of compositional asymmetry (viz., either $x_A > x_B$ or $x_A < x_B$), but gradually fades away when $x_A \sim x_B$.

### 3.2. Isothermal Crystallization

To understand the effect of thermal history (and cooling pathway) on crystallization and morphological development, we have carried out isothermal crystallization. We first equilibrate the sample system at $U_p = 0$ (athermal state). Following this, we quench the sample system to $U_p = 0.6$, annealed for $10^5$ Monte Carlo steps (*MCS*) and measure $X_A$, $X_B$ and $X$ as a function of *MCS*. Figure 12a represents the evolution of overall crystallinity with *MCS* for all the $x_B$ values investigated ($\lambda = 1$). We also calculate scaled crystallinity, $X_c^* = \left( X_c - X_c^i \right) / \left( X_c^f - X_c^i \right)$, ranges from 0 to 1.0 as a function of *MCS* for all the compositions investigated (Figure 12b). $X_c^i$ represents the crystallinity at the beginning of the isothermal experiment, and $X_c^f$ represents the crystallinity at the end of isothermal annealing (viz., $X_c$ at the end of $1 \times 10^5$ *MCS*). Figure 13a and b represents the evolution of overall crystallinity and scaled crystallinity for $\lambda = 6$. The trend in overall crystallinity as a function of $x_B$ reveals that the kinetic pathway for crystallization of two crystallisable components is a strong function of the relative composition. The overall saturated crystallinity shows a non-monotonic trend with $x_B$ which is attributed to the compositional heterogeneity (Figure 14). When either of the components is large enough, the overall crystallinity is dominated by that component compared to the other. When $x_A \sim x_B$, both the components compete with each other and due to the interface rigidity the crystal growth is restricted. This non-monotonic trend in crystallinity is equally observed in both weak and strong segregations, although the level of crystallinity is drastically less with a wider variation across the composition in case of strong segregation, which is attributed to the strong repulsive interaction between A- and B-polymer making the interface more rigid and influencing the crystal development.[6]

To get an estimate of the effect of composition on rate of crystallization, we calculate the half-time (in terms of number of $MCS$) of crystallization for all the compositions at $\lambda = 1$ and 6, and plotted in Figure 15a and 15b, respectively. Higher the vale of $t_{1/2}$, slower is the rate of crystallization (viz., rate of crystallization $\sim t_{1/2}^{-1}$). The results at $\lambda = 1$ shows a non-monotonic trend of crystallization half-time with composition, $x_B$. At high enough composition of $x_B$, the rate of crystallization of B-polymer is faster than that of A-polymer, due to the presence of large number of B-units. As discussed before, competitive crystallization at $x_A \sim x_B$ leads to the formation of less crystalline materials, observed also in the trend of overall saturation crystallinity (Figure 14). On the other hand, at strong segregation strength (viz., $\lambda = 6$), the rate of crystallization is extremely slow due to the strong repulsive interaction between the components. Similar non-monotonic trend is also observed with extremely slow rate of crystallization of individual components when it is less in proportion. However, the saturation crystallinities of both the components are extremely small due to the strong segregation strength, and follow a non-monotonic trend (see Table S1, Supplementary information).[40]

Polymer crystallization is not truly an equilibrium phenomenon. The development of crystalline structure largely depends on crystallization temperature as well as on cooling pathway. In our present study on binary polymer melt, we model A- and B- polymer with different crystallizability, and we expect that the quench depth may influence crystallization and morphological development. We carried out isothermal experiment in two steps as follows: in the first step we equilibrate the sample system at $U_p = 0$ and quench to $U_p = 0.28$, annealed for $10^5$ $MCS$; and in the second step, we quench to $U_p = 0.6$ from $U_p = 0.28$, annealed for $10^5$ $MCS$. At $U_p = 0.28$ (temperature above the melting point of B-polymer), A-polymer crystallizes, while B-polymers are still in a molten state. Upon quenching to $U_p = 0.6$ (from $U_p = 0.28$), B-polymer starts to crystallize within the domains created during macrophase separation followed by A-polymer crystallization. Table 2 shows that the crystallinity of A-polymer is significantly enhanced during two-step crystallization, compared to one-step ($\lambda = 1$). We interpret this enhancement in crystallinity as follows. When we quench the sample system to $U_p = 0.28$, temperature below the melting point of A-polymer, but above the melting point of B-polymer, A-polymer crystallizes without much hindrance from B-polymer, since B-polymer is in a molten state. Next stage, when quench to $U_p = 0.6$, below the melting point of B-polymer (also A-polymer), crystallization of B-polymer does not experience hindrance from A-polymer, since A-polymer is already crystallized within its own domains. Thus, the mode of crystallization in case of two-step cooling is sequential crystallization, as evident by experimental work on blends of PHB/PLLA,[24] where PLLA crystallizes at 120 °C followed by PHB at 90 °C. Similarly, in PBS/PEO[9] blend, crystallization of PBS (at 95 °C) is followed by the crystallization of PEO at 50 °C; and also in PLLA/PEO[46] blend, wherein PLLA crystallizes at 125 °C followed by PEO at 95 °C. In one-step isothermal crystallization, when we quench to $U_p = 0.6$, the crystallization temperature is well below the melting point of both A- and B-polymer, and hence the driving force for the crystallization of both the polymers is adequate. In our model, the crystallization driving force for A-polymer is higher than B-polymer; as a result, A-polymer possesses higher crystallinity than B-polymer as the crystallization progresses with Monte Carlo steps. Thus, the mode of crystallization is referred to as coincident crystallization, where both the polymers compete with each other to crystallize simultaneously. This observation is in close agreement with the experimental work on PHB/PLLA[24] blend, crystallized at 110 °C. It has been observed that although PLLA and PHB crystallize simultaneously, the rate of crystallization of PLLA is significantly faster than that of PHB. However, in the blend of PBS and PEO,[9] during one-step cooling at 50 °C, the exceptionally fast rate of crystallization of PBS filled the entire space before the crystallization of PEO starts. Therefore, the mechanism of crystallization becomes sequential. In one-step isothermal crystallization (viz., quenching to $U_p = 0.6$), inter-block entanglement restricts the development of crystallinity of A-polymer. However, the magnitude of saturation crystallinity of B-polymer in both the experiments is close to each other. The saturation crystallinity of B-polymer is 0.65 and 0.68 (Table 3) for one-step and two-step isothermal crystallization, respectively. Thus, two-step crystallization yields better crystalline structure for A-polymer over B-polymer. Analysis based on lamellar thickness reveals that the lamellar thickness of A-polymer at $U_p = 0.6$ (one-step quenching) is less than that of $U_p = 0.28$, during two-step cooling (Table S2, Supplementary information[40]), which is in accord with the Hoffman-Weeks formulation[47] and recent experimental observation on the blends of PBS/PEO.[9] However, lamellar thickness of B-polymer, $\langle l_B \rangle$ at $U_p = 0.6$ in one-step cooling appears to be almost identical to that of at $U_p = 0.6$, in two-step cooling, for all the compositions investigated (Table S3, Supplementary information). This non-intuitive trend in lamellar thickness may be interpreted as follows. Lamellar thickness is largely governed by the degree of undercooling $\left( \Delta T = T_m^\alpha - T \right)$. The effective $\Delta T$ for B-polymer remains same in both one- and two-step isothermal processes, when quenched to $U_p = 0.6$. Therefore, the development of crystal thickness remains identical in both the cases. Higher crystallinity results due to the crystal growth along the lateral direction. Table 2 and 3 also show the variation in crystallinity with composition in one- and two-step cooling, at $\lambda = 1$. Comparisons of saturation crystallinity of A- and B-polymer at $\lambda = 6$ are available in Table S4 and Table S5 of supplementary information,[40] respectively. At $U_p = 0.28$, in two-step cooling, as $x_B$ increases, crystallinity of A-polymer remains almost constant except at very high $x_B$ (0.875), where it significantly decreases (Table 2). Similar decreasing trend is also observed at $U_p = 0.6$, in two-step cooling. This is attributed to the less proportion of A-polymer compared to B-polymer. On the other hand, crystallinity of B-polymer increases at a higher value of $x_B$ ($U_p = 0.6$, $x_B = 0.875$), due to the presence of large proportion of the B-units. However, in every composition, two-step cooling yields higher crystallinity than one-step cooling. Figure 16 and 17 display the snapshots of one- and two-step isothermal crystallization at $\lambda = 1$, $x_B = 0.25$ and 0.75, respectively. At strong segregation strength (viz., $\lambda = 6$), the values of crystallinity of both the polymers follow a similar trend as $\lambda = 1$, but with extremely less crystallinity. In one-step cooling, most of the sample produce amorphous structure. Two-step in comparison to one-step cooling yields much higher crystallinity for all the compositions investigated.

## 4. Conclusions

In this paper, we present simulation results on crystallization of binary polymer blends to elucidate the effect of composition heterogeneity on crystallization and morphological development. In our model A- and B-polymers become immiscible as we cool the sample system from a high temperature melt. The chemical dissimilarity between A- and B-polymer leads to a repulsive interaction, manifested by segregation strength. As we start cooling the sample system, they phase separate via macrophase separation prior to crystallization. We model A-polymer as high melting polymer, and as a result crystallization of A-polymer precedes the crystallization of B-polymer. . During macrophase separation, A- and B-polymers form their respective domains and subsequent crystallization happens with the domains created during macrophse separation. In weakly segregated systems (viz., $\lambda = 1$), with increasing compositional asymmetry in the blends (viz., one component is present in higher proportion than the other), we observe a non-monotonic trend in transition temperature, crystallization temperature, mean square radius of gyration and also lamellar thickness. As the composition ~ 1:1, the respective values tend to possess a lowest one, which is attributed to the competition between two components in forming individual domains. Similar trend is also observed with strongly segregated systems (viz., $\lambda = 6$), except for the value of $\left\langle R_g^2 \right\rangle$ which shows an opposite trend. As the composition approaches ~ 1:1, due to the competition between two components, large number of inter-connected smaller size domains are formed, which give rise to a higher value of $\left\langle R_g^2 \right\rangle$. As the composition becomes more asymmetric in nature, the major component excludes the minor component from its domains and as a result, the value of $\left\langle R_g^2 \right\rangle$ again decreases. At high enough value of $x_B$ (0.75 and 0.875), lamellar thickness of A-polymer increases, which is attributed to the dilution effect shown by B-polymer. When composition of A-polymer is very less in the system, B-polymer acts like a "solvent" and reduces the topological restriction favoring the formation of thicker crystals. At weak segregation strength, dilution effect dominates to give a non-monotonic trend in the crystallization behavior. However, at strong segregation strength, the repulsive strength between two polymers coupled with the dilution effect, shows a dramatic change in the crystallization behavior. Isothermal crystallization clearly shows that the crystallization behavior is strongly influenced by the relative composition of the blend for both the segregation strengths (viz., $\lambda = 1$ and 6). The rate of crystallization measured in terms of half-time of crystallization also shows a strong compositional dependency. We have also elucidated the path dependent crystallization behavior. Sequential crystallization in two-step cooling yields higher crystallinity with thicker crystals compared to coincident crystallization in one-step cooling, for all the compositions investigated.

**Acknowledgement:** Computational facility supported by the SERB, Department of Science and Technology (DST), Government of India (sanction letter no. SR/S3/CE/0069/2010) is highly acknowledged.

**Table Captions:**

**Table 1 Comparison in saturation fractional crystallinity of A- ( $X_A$ ) and B-polymer ( $X_B$ ) at $\lambda$ = 1 and 6, non-isothermal crystallization.**

**Table 2 Comparison in fractional crystallinity of A-polymer ( $X_A$ ) at $\lambda$ = 1 during one-step and two-step isothermal crystallization.**

**Table 3 Comparison in fractional crystallinity of B-polymer ( $X_B$ ) at $\lambda$ = 1 during one-step and two-step isothermal crystallization.**

**Figure Captions:**

**Figure 1 Snapshot of the simulation system representing a homogeneous melt of binary polymer blend at $U_p$ = 0, composition ( $x_B$ ) = (a) 0.25 and (b) 0.75. Blue and magenta line represents segments of A- and B-polymer, respectively.**

**Figure 2 Change in specific heat, $C_v$ with $U_p$ for different compositions, at (a) $\lambda$ = 1 and (b) $\lambda$ = 6. Inset of the graph shows change in transition point, $U_p^*$ with $x_B$ for A- and B-polymer. The lines joining the points are meant only as a guide to the eye.**

**Figure 3 Change in mean square displacement of center of mass ( $d_{cm}^2$ ) of A- and B-polymer with composition, $x_B$ at $U_p$ = 0.28 for (a) $\lambda$ = 1 and (b) $\lambda$ = 6. The lines joining the points are meant only as a guide to the eye.**

**Figure 4** Change in macrophase separation point, $U_p^{\#}$ (calculated from $C_{v\_AB}$ vs. $U_p$) with $x_B$ for **(a)** $\lambda = 1$ and **(b)** $\lambda = 6$. **The lines joining the points are meant only as a guide to the eye.**

**Figure 5** Snapshots of macrophase separated melt at $U_p = 0.1$ for **(a)** $\lambda = 1$, $x_B = 0.25$, **(b)** $\lambda = 6$, $x_B = 0.25$, **(c)** $\lambda = 1$, $x_B = 0.75$ and **(d)** $\lambda = 6$, $x_B = 0.75$ . **Blue and magenta line represents segments of A- and B-polymer, respectively.**

**Figure 6** Change in mean square radius of gyration, $\langle R_g^2 \rangle$ with $U_p$ for different compositions at **(a)** $\lambda = 1$ and **(b)** $\lambda = 6$. **The insets show the change in** $\langle R_g^2 \rangle$ **with composition at** $U_p = 0.28$ and $0.6$. **The lines joining the points are meant only as a guide to the eye.**

**Figure 7 (a)** Change in overall crystallinity with $U_p$ *(~1/T)* for all the compositions (including pure A-polymer, $x_B = 0$), at $\lambda = 1$. **(b)** Change in saturated crystallinity with compositions for A-polymer, B-polymer and overall at $\lambda = 1$. **The lines joining the points are meant only as a guide to the eye.**

**Figure 8 (a)** Change in overall crystallinity with $U_p$ *(~1/T)* for all the compositions (including pure A-polymer, $x_B = 0$), at $\lambda = 6$. **(b)** Change in saturated crystallinity with compositions for A-polymer, B-polymer and overall at $\lambda = 6$. **The lines joining the points are meant only as a guide to the eye.**

**Figure 9** Snapshots of semi-crystalline structure at $U_p = 0.6$ for **(a)** $x_B = 0.25$, $\lambda = 1$, **(b)** $x_B = 0.25$, $\lambda = 6$, **(c)** $x_B = 0.75$, $\lambda = 1$ and **(d)** $x_B = 0.75$, $\lambda = 6$. **Blue lines represent crystalline bonds of A-polymer, magenta lines represent crystalline bonds of B-polymer and yellow lines represent non-crystalline bonds of both the polymers.**

**Figure 10** Change in average crystallite size, $\langle S \rangle$ of A- and B-polymer with $x_B$ at **(a)** $\lambda = 1$ and **(b)** $\lambda = 6$. **The lines joining the points are meant only as a guide to the eye.**

**Figure 11 Change in average lamellar thickness, $\langle l \rangle$ of A- and B-polymer with $x_B$ at (a) $\lambda = 1$ and (b) $\lambda = 6$. The lines joining the points are meant only as a guide to the eye.**

**Figure 12 Change in isothermal (a) Overall crystallinity and (b) Scaled crystallinity with number of Monte Carlo steps (*MCS*) for different compositions, at $\lambda = 1$. The lines joining the points are meant only as a guide to the eye.**

**Figure 13 Change in isothermal (a) Overall crystallinity and (b) Scaled crystallinity with number of Monte Carlo steps (*MCS*) for different compositions, at $\lambda = 6$. The lines joining the points are meant only as a guide to the eye.**

**Figure 14 Change in saturated overall crystallinity in isothermal crystallization with $x_B$, at $\lambda = 1$ and 6. The lines joining the points are meant only as a guide to the eye.**

**Figure 15 Change in crystallization half-time with composition, xB for A- and B-polyemr at (a) $\lambda = 1$ and (b) $\lambda = 6$. The lines joining the points are meant only as a guide to the eye.**

**Figure 16. Snapshots of semi-crystalline structures for $\lambda = 1$, $x_B = 0.25$ (a) at $U_p = 0.6$ (one-step isothermal cooling)., (b) at $U_p = 0.28$ (during two-step isothermal cooling), (c) at $U_p = 0.6$ (during two-step isothermal cooling). Blue and magenta lines represent crystalline bonds of A- and B-polymer, respectively; yellow lines represent non-crystalline bonds of both the polymers.**

**Figure 17. Snapshots of semi-crystalline structures for $\lambda = 1$, $x_B = 0.75$ (a) at $U_p = 0.6$ (one-step isothermal cooling)., (b) at $U_p = 0.28$ (during two-step isothermal cooling), (c) at $U_p = 0.6$ (during two-step isothermal cooling). Blue and magenta lines represent crystalline bonds of A- and B-polymer, respectively; yellow lines represent non-crystalline bonds of both the polymers.**

**Table 1**

| | Weak segregation, $\lambda = 1$ | | Strong segregation, $\lambda = 6$ | |
|---|---|---|---|---|
| **Composition ($x_B$)** | $X_A$ | $X_B$ | $X_A$ | $X_B$ |
| **0.125** | 0.763 | 0.748 | 0.680 | 0.476 |
| **0.25** | 0.749 | 0.735 | 0.617 | 0.482 |
| **0.375** | 0.751 | 0.743 | 0.512 | 0.341 |
| **0.5** | 0.745 | 0.742 | 0.510 | 0.350 |
| **0.625** | 0.750 | 0.735 | 0.434 | 0.343 |
| **0.75** | 0.732 | 0.754 | 0.564 | 0.564 |
| **0.875** | 0.735 | 0.766 | 0.622 | 0.687 |

**Table 2**

|  | Two-step cooling | | One-step cooling |
|---|---|---|---|
| Composition ($x_B$) | $U_p = 0.28$ | $U_p = 0.6$ | $U_p = 0.6$ |
| 0.125 | 0.747 | 0.776 | 0.683 |
| 0.25 | 0.738 | 0.765 | 0.664 |
| 0.375 | 0.726 | 0.752 | 0.660 |
| 0.5 | 0.727 | 0.750 | 0.653 |
| 0.625 | 0.719 | 0.746 | 0.645 |
| 0.75 | 0.715 | 0.752 | 0.645 |
| 0.875 | 0.673 | 0.736 | 0.641 |

**Table 3**

| Composition ($x_B$) | Two-step cooling | | One-step cooling |
|---|---|---|---|
| | $U_p = 0.28$ | $U_p = 0.6$ | $U_p = 0.6$ |
| 0.125 | 0.085 | 0.684 | 0.632 |
| 0.25 | 0.097 | 0.673 | 0.640 |
| 0.375 | 0.103 | 0.683 | 0.640 |
| 0.5 | 0.107 | 0.685 | 0.641 |
| 0.625 | 0.111 | 0.683 | 0.648 |
| 0.75 | 0.112 | 0.694 | 0.663 |
| 0.875 | 0.112 | 0.718 | 0.690 |

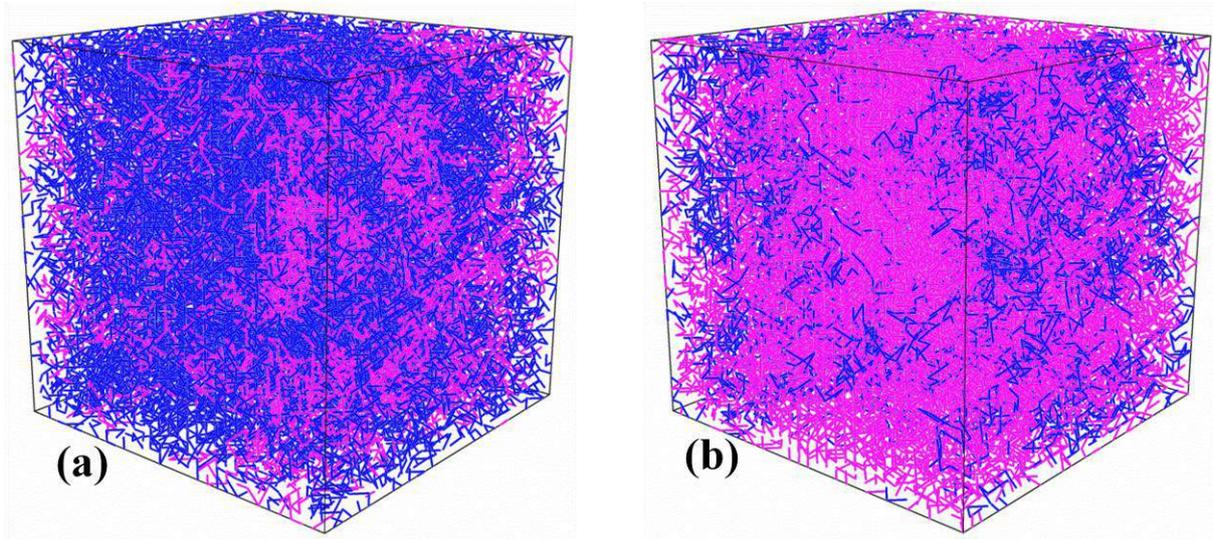

**(a)**  **(b)**

**Figure 1**

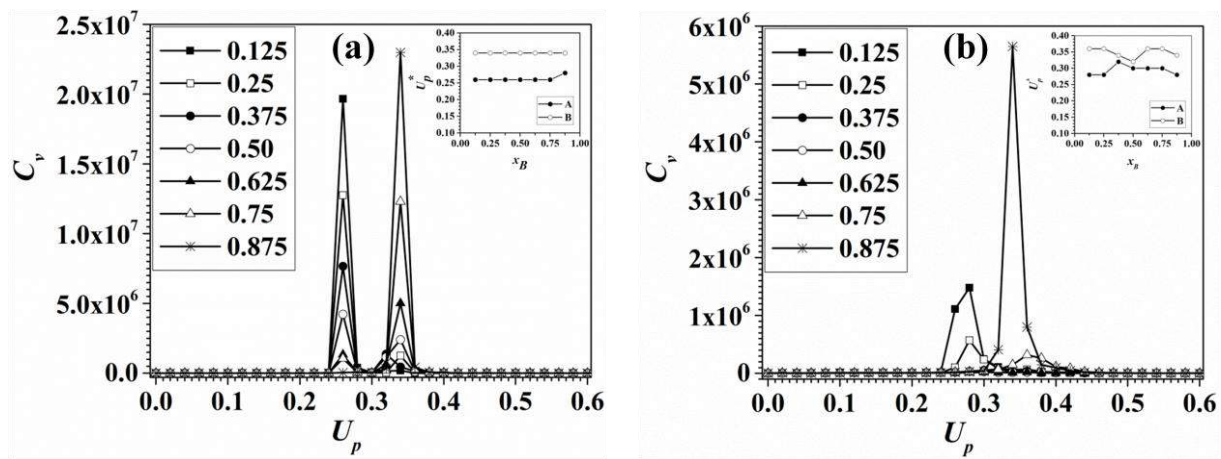

**Figure 2**

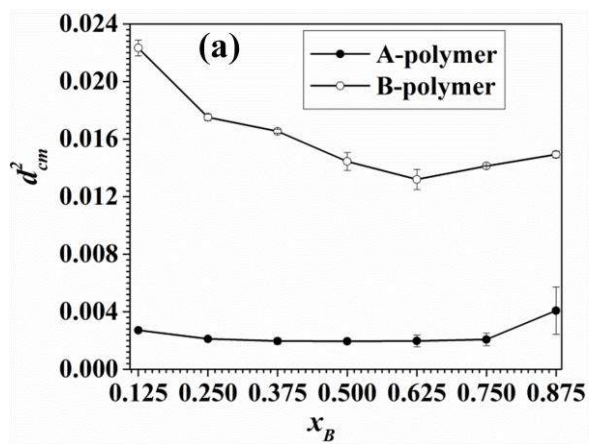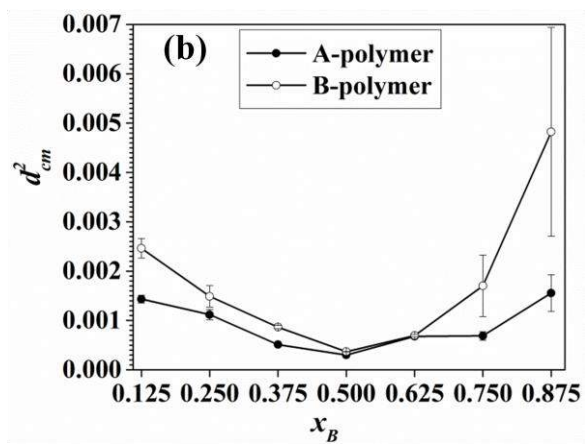

**Figure 3**

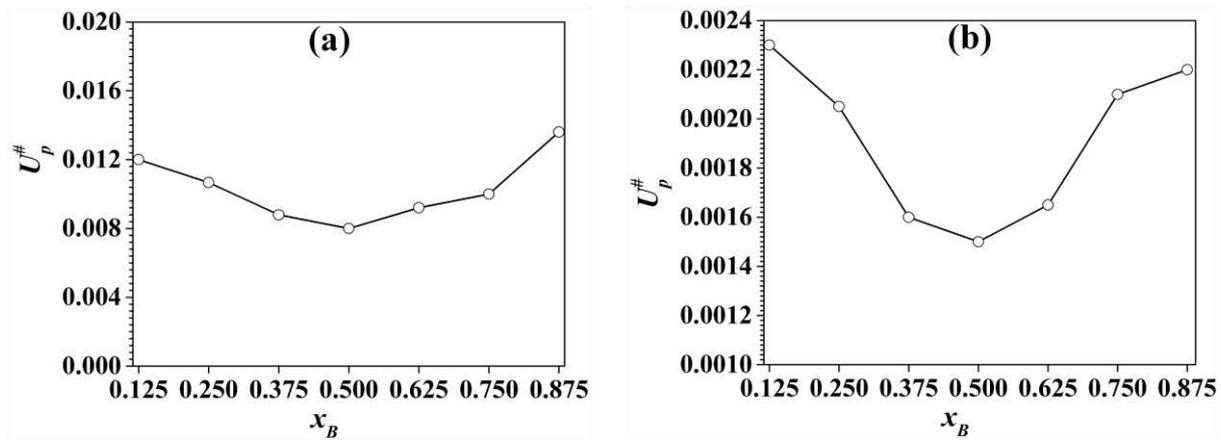

**Figure 4**

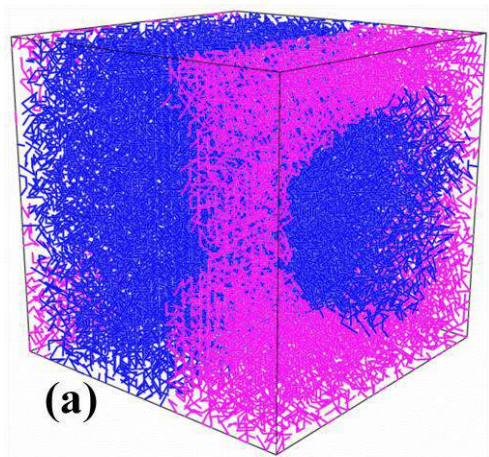 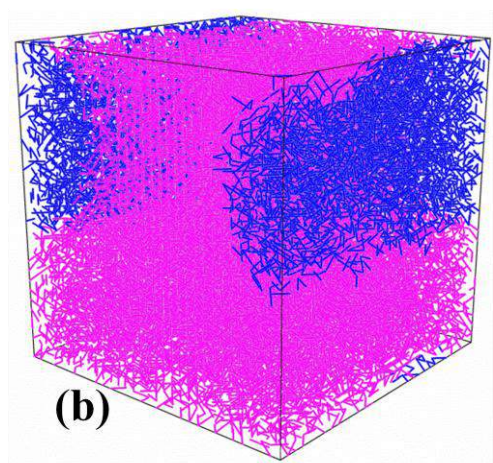

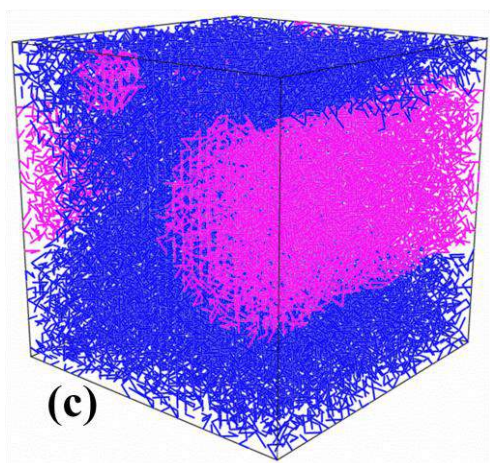 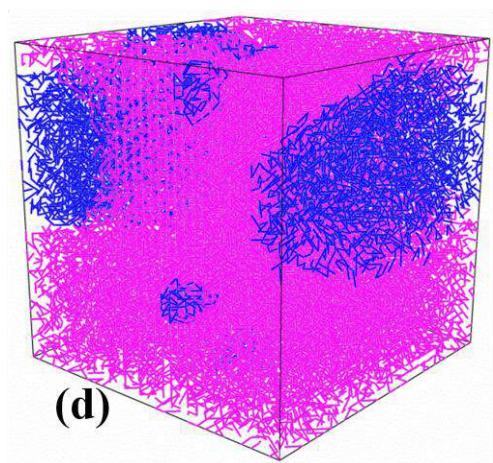

**Figure 5**

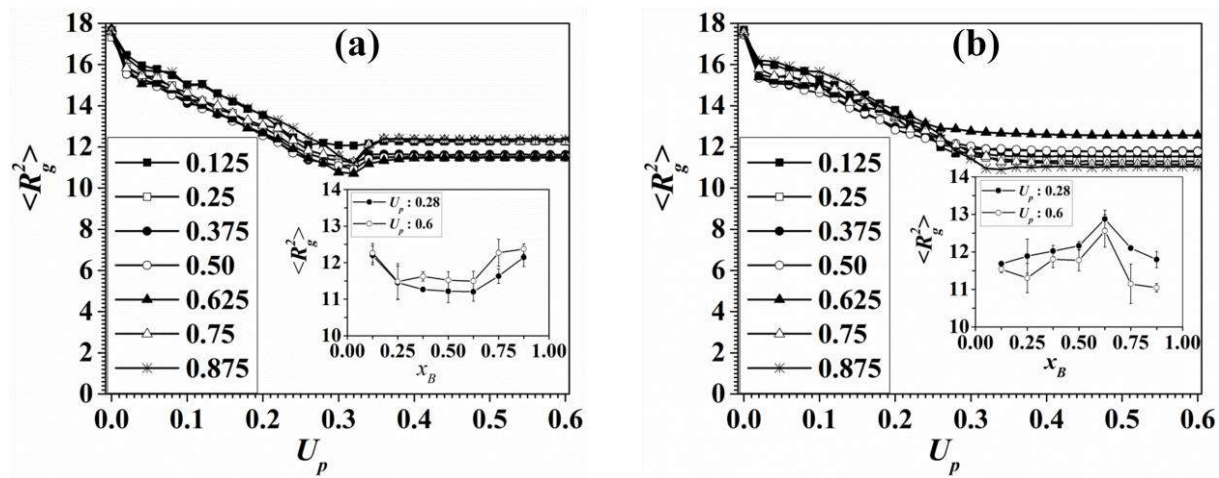

**Figure 6**

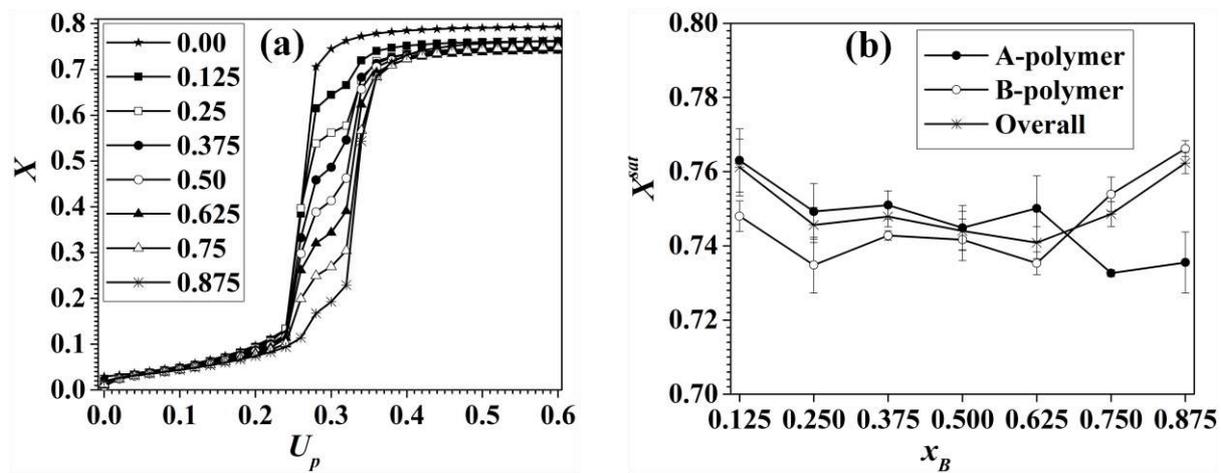

**Figure 7**

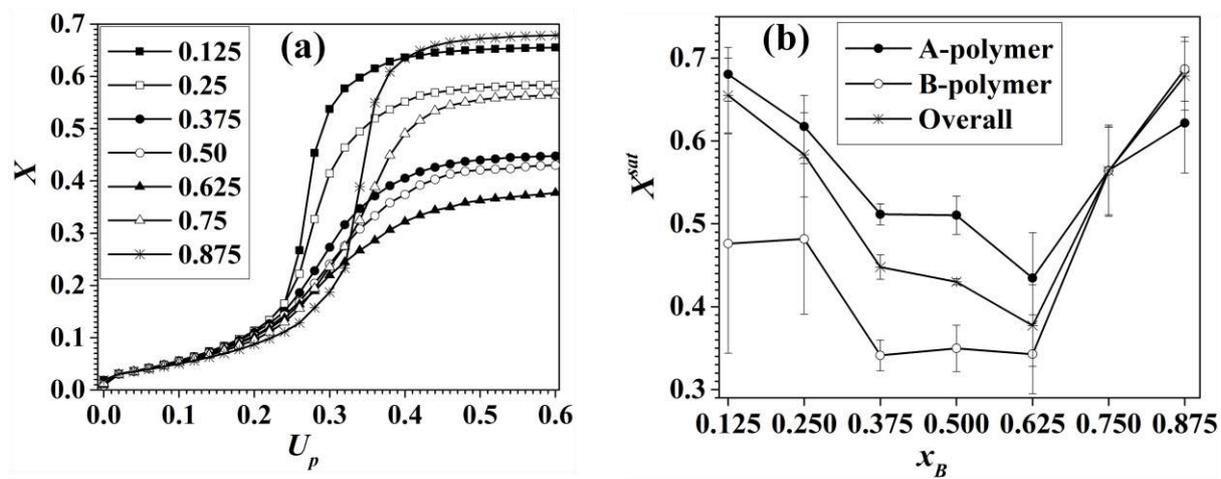

**Figure 8**

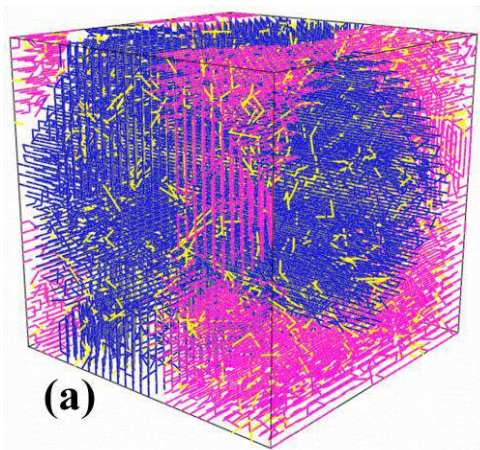 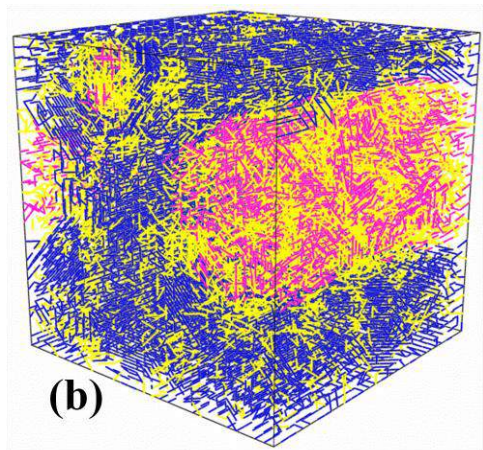

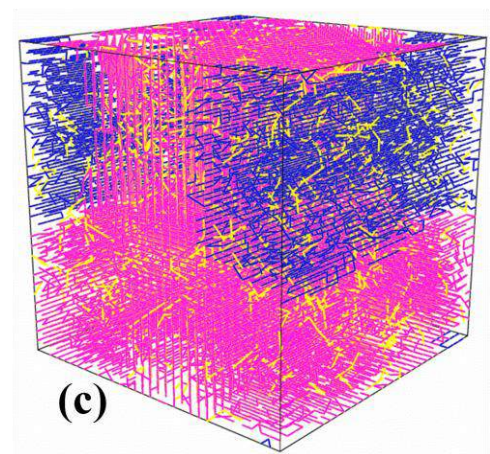 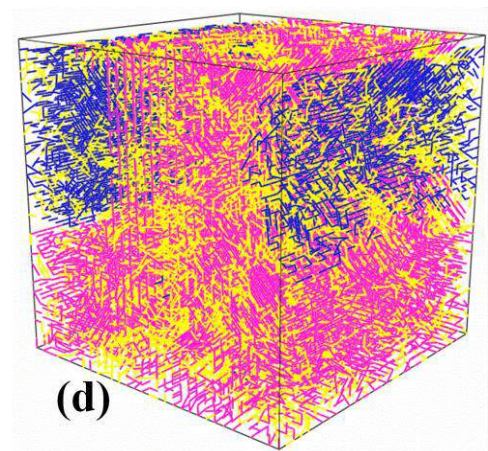

**Figure 9**

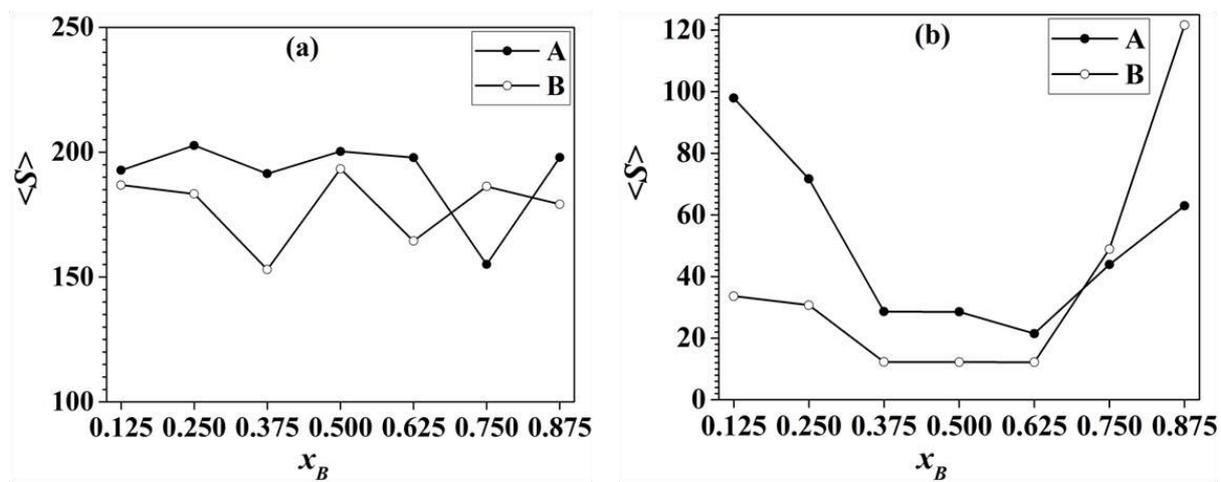

**Figure 10**

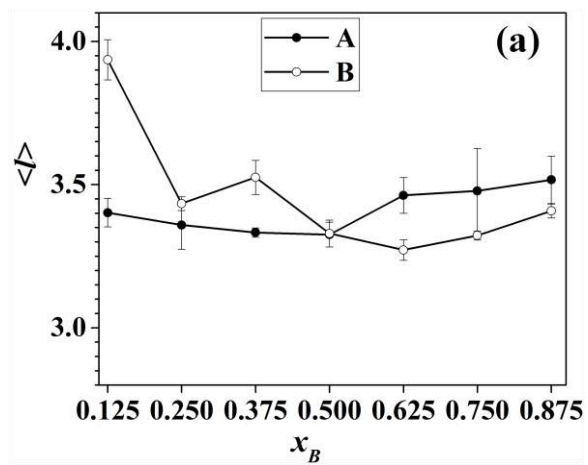 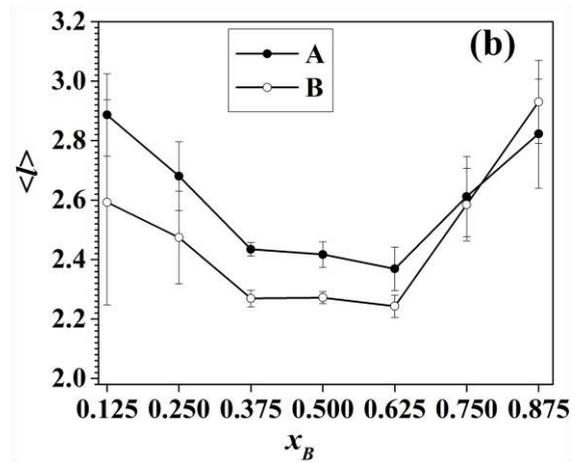

**Figure 11**

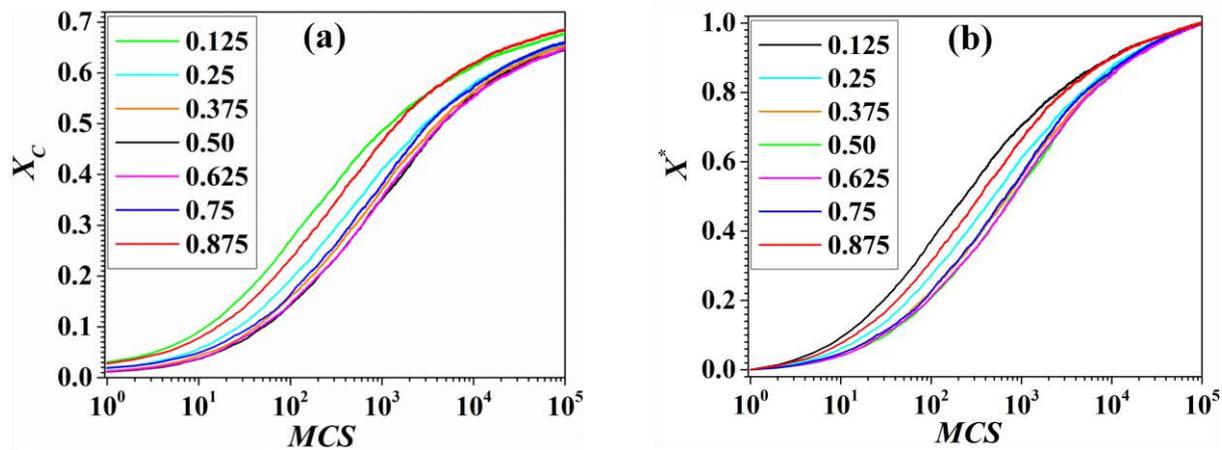

**Figure 12**

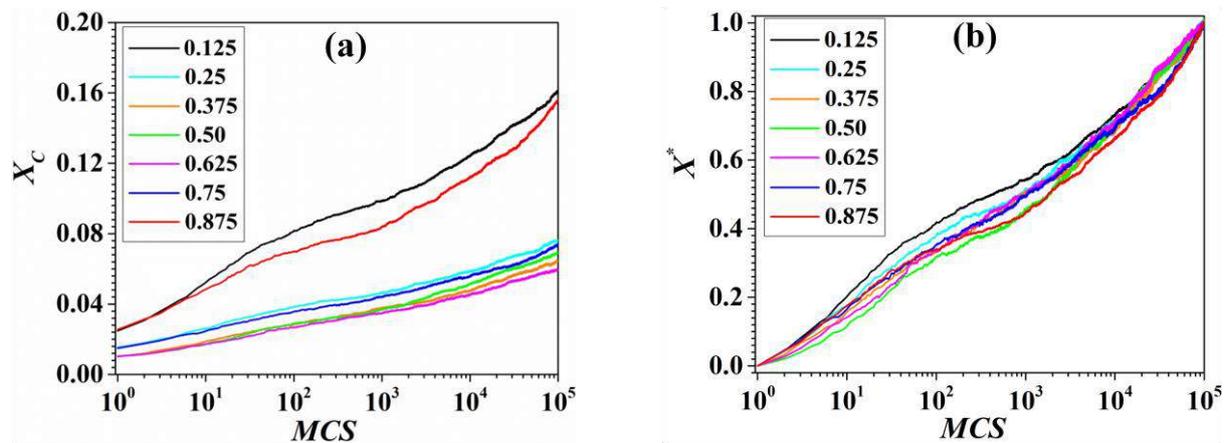

**Figure 13**

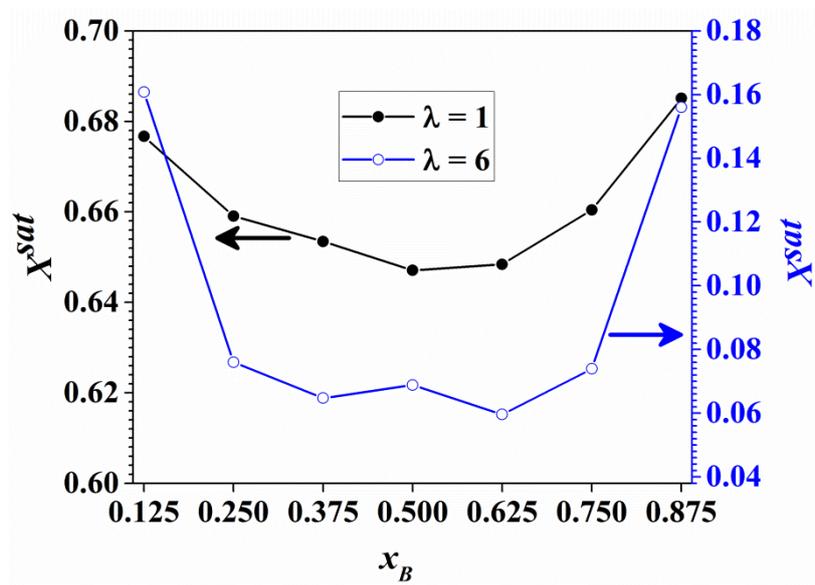

**Figure 14**

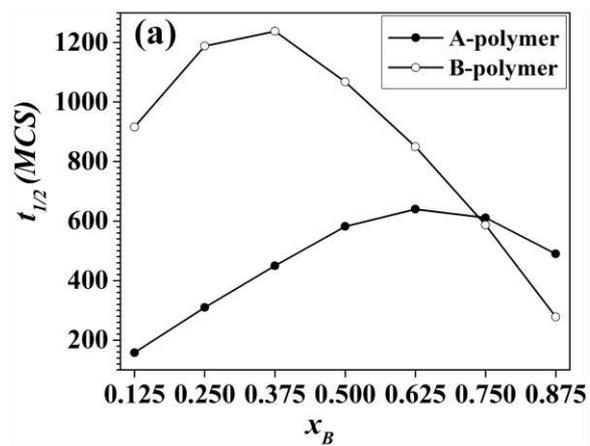
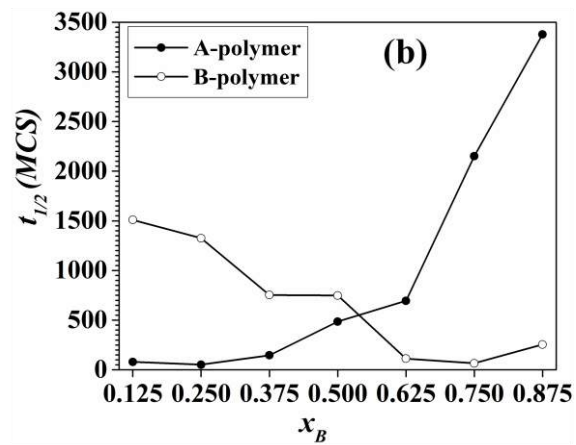

**Figure 15**

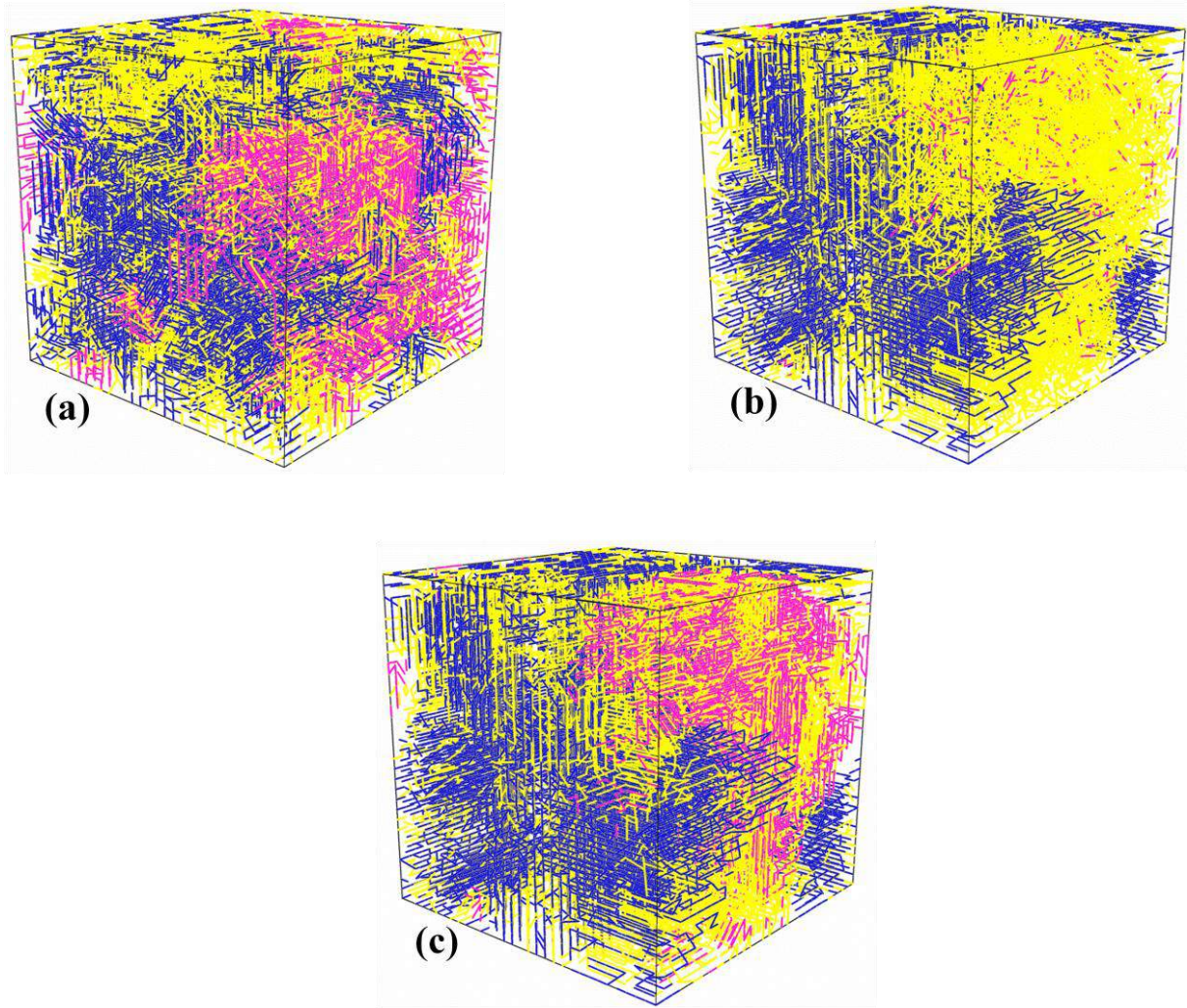

**Figure 16**

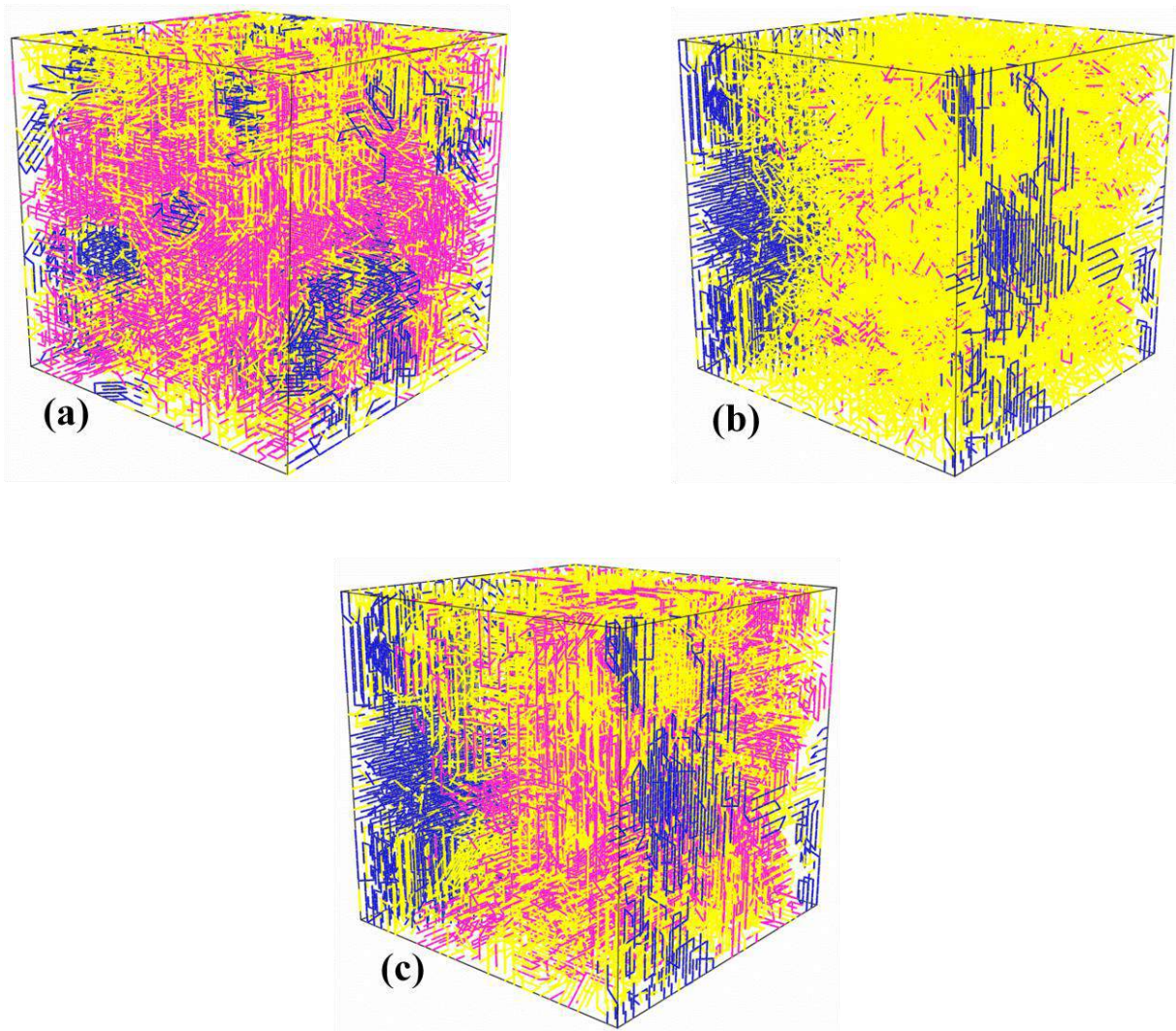

**Figure 17**

**Supplementary Information**

**Composition Heterogeneity Induced Crystallization in Double Crystalline Binary Polymer Blends**

Ashok Kumar Dasmahapatra[*]

Department of Chemical Engineering, Indian Institute of Technology Guwahati, Guwahati - 781039, Assam, India

**Fig. S1 Snapshots at $U_p = 0$ for various compositions, $x_B$: (a) 0.125, (b) 0.375, (c) 0.5, (d) 0.625 and (e) 0.875. Blue and magenta colors represent bonds of A- and B-polymers, respectively.**

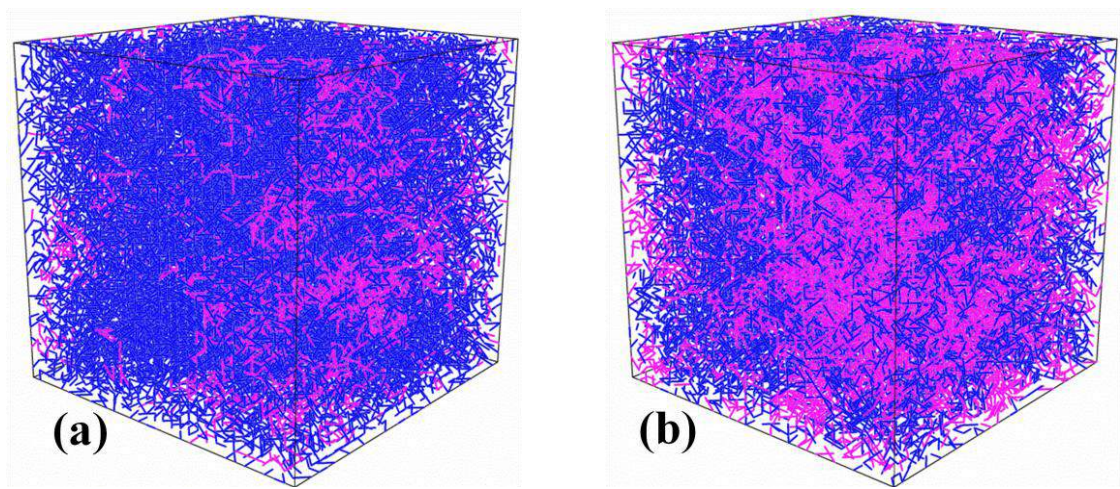

[*] Corresponding author: Phone: +91-361-2582273, Fax: +91-361-2582291, Electronic mail: akdm@iitg.ernet.in

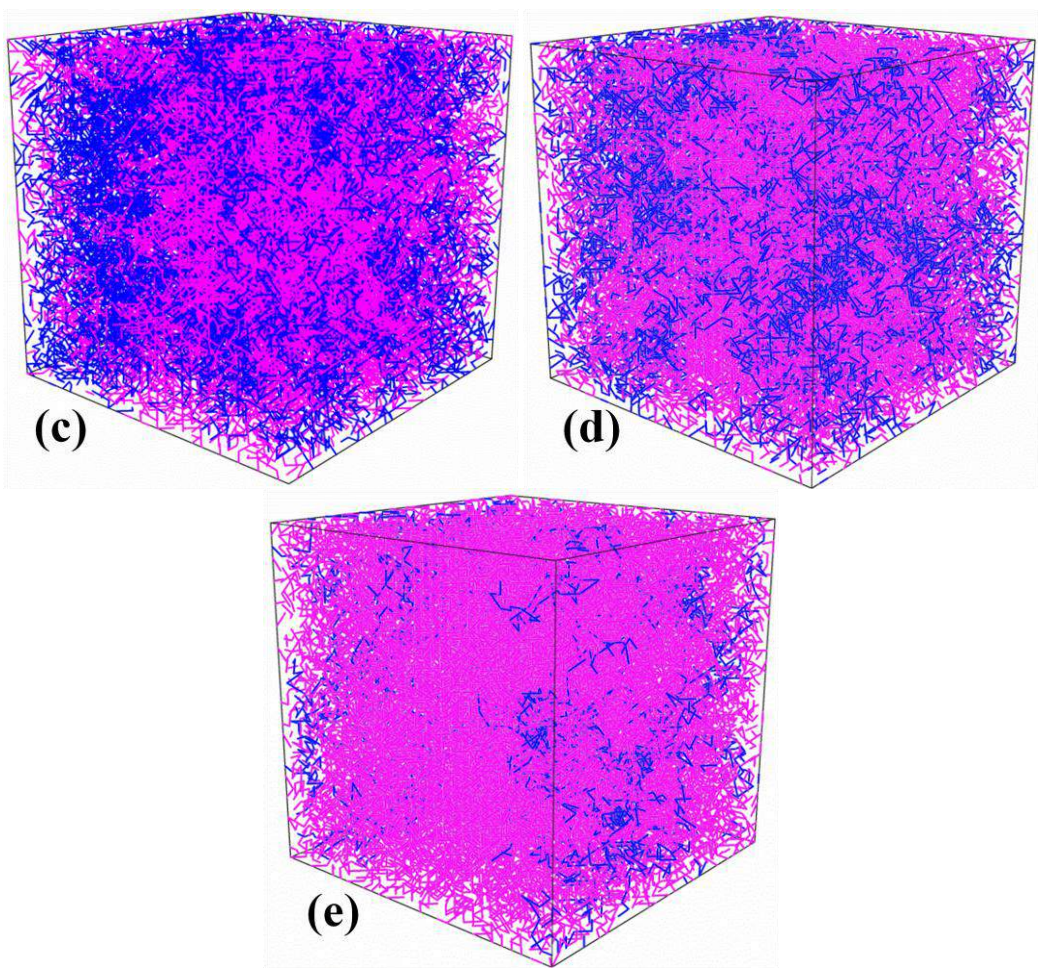

**Fig. S2** Snapshots of macrophase separated melt at $U_p$ = 0.1 during non-isothermal crystallization, for various compositions and $\lambda$: (a) $\lambda$ = 1, $x_B$ = 0.125, (b) $\lambda$ = 1, $x_B$ = 0.375, (c) $\lambda$ = 1, $x_B$ = 0.5, (d) $\lambda$ = 1, $x_B$ = 0.625, (e) $\lambda$ = 1, $x_B$ = 0.875, (f) $\lambda$ = 6, $x_B$ = 0.125, (g) $\lambda$ = 6, $x_B$ = 0.375, (h) $\lambda$ = 6, $x_B$ = 0.5, (i) $\lambda$ = 6, $x_B$ = 0.625 and (j) $\lambda$ = 6, $x_B$ = 0.875. Blue and magenta colors represent bonds of A- and B-polymers, respectively.

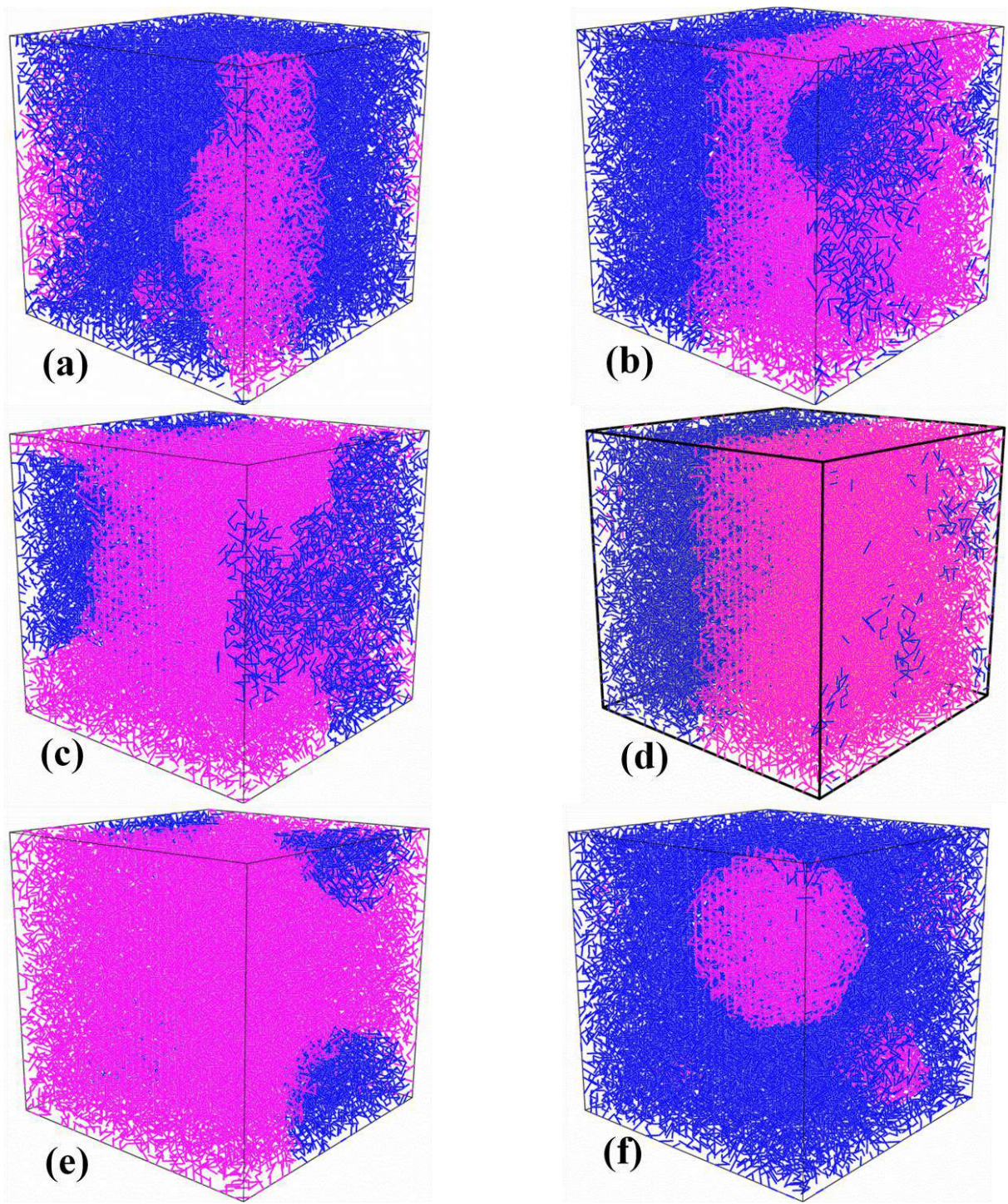

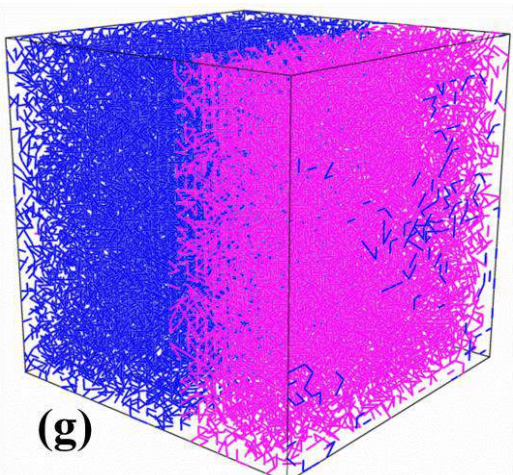
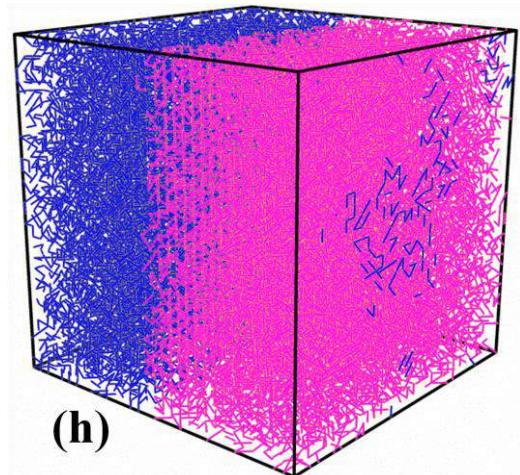

**(g)** **(h)**

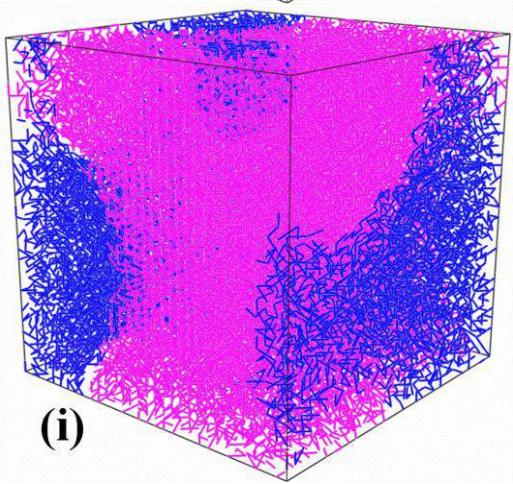
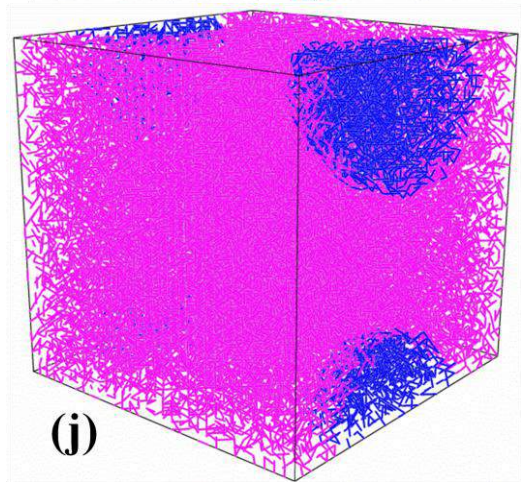

**(i)** **(j)**

**Fig. S3** Snapshots of crystalline structure at $U_p$ = **0.28 during non-isothermal crystallization for various compositions at** $\lambda = 1$: **(a)** $x_B = 0.125$, **(b)** $x_B = 0.25$, **(c)** $x_B = 0.375$, **(d)** $x_B = 0.50$, **(e)** $x_B = 0.625$, **(f)** $x_B = 0.75$, **(g)** $x_B = 0.875$. **Blue and magenta colors represent crystalline bonds of A- and B-polymers, respectively. Yellow color represents non-crystalline bonds of both the polymers.**

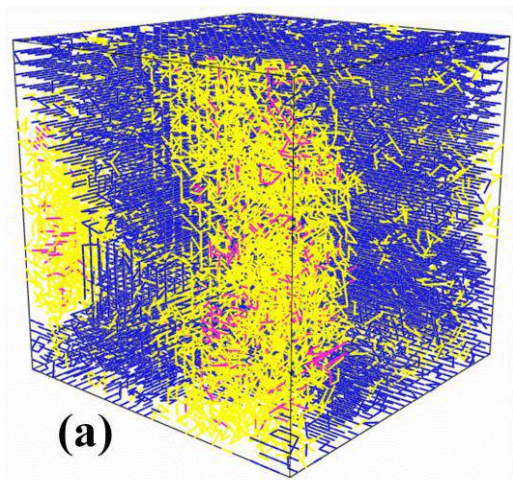

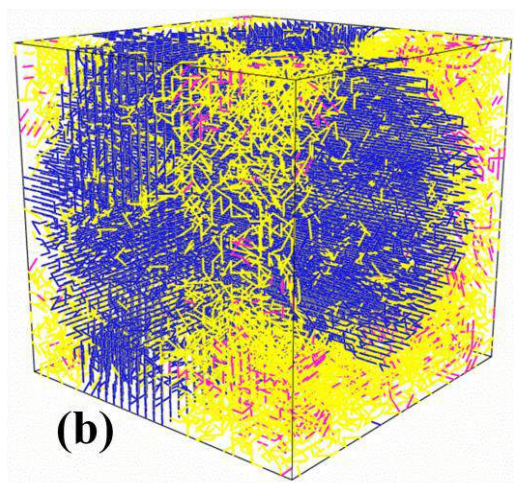

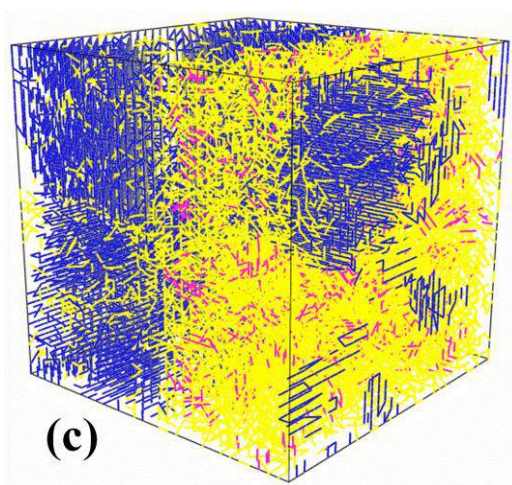

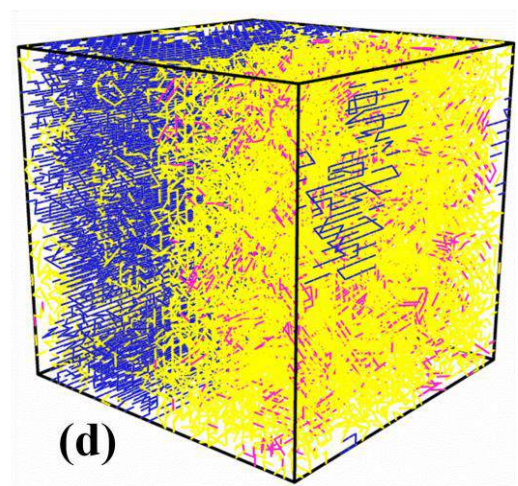

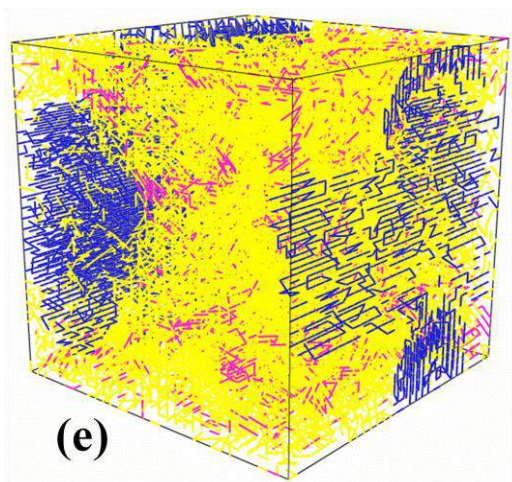
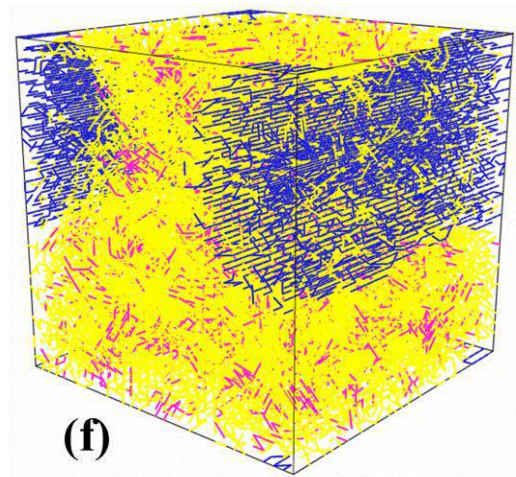
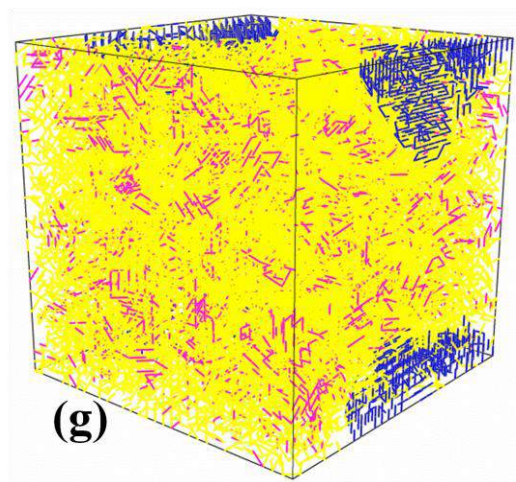

**Fig. S4** Snapshots of crystalline structure at $U_p$ = 0.6 during non-isothermal crystallization, for various compositions and $\lambda$: (a) $\lambda$ = 1, $x_B$ = 0.125, (b) $\lambda$ = 1, $x_B$ = 0.375, (c) $\lambda$ = 1, $x_B$ = 0.5, (d) $\lambda$ = 1, $x_B$ = 0.625, (e) $\lambda$ = 1, $x_B$ = 0.875, (f) $\lambda$ = 6, $x_B$ = 0.125, (g) $\lambda$ = 6, $x_B$ = 0.375, (h) $\lambda$ = 6, $x_B$ = 0.5, (i) $\lambda$ = 6, $x_B$ = 0.625 and (j) $\lambda$ = 6, $x_B$ = 0.875. Blue and magenta colors represent crystalline bonds of A- and B-polymers, respectively. Yellow color represents non-crystalline bonds of both the polymers.

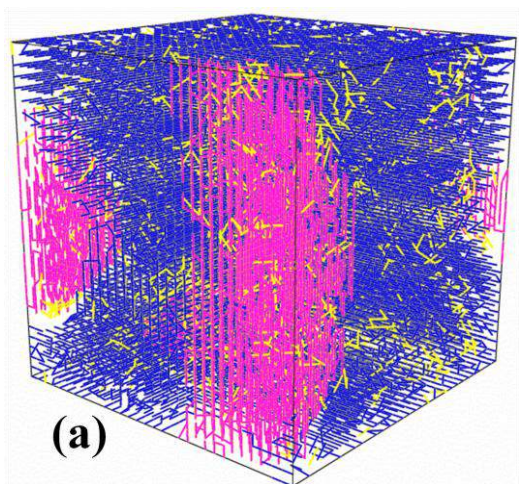

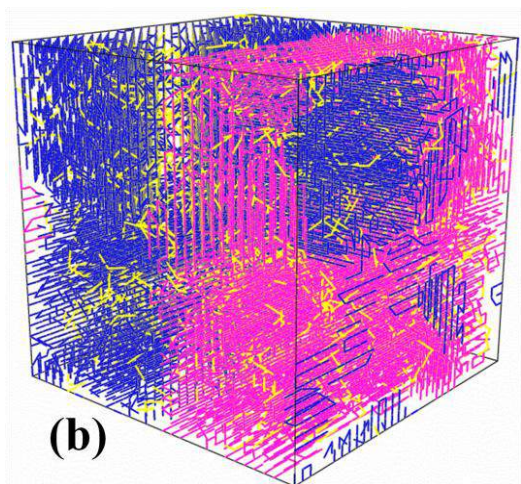

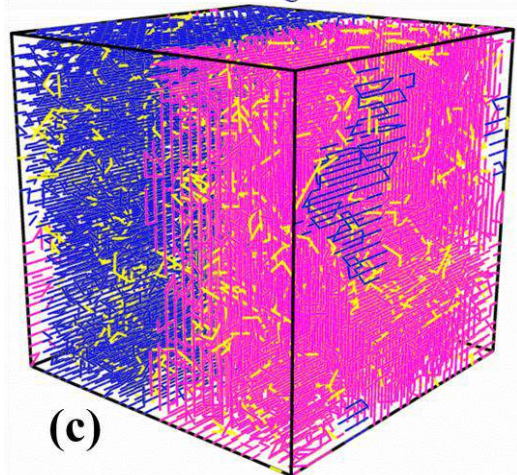

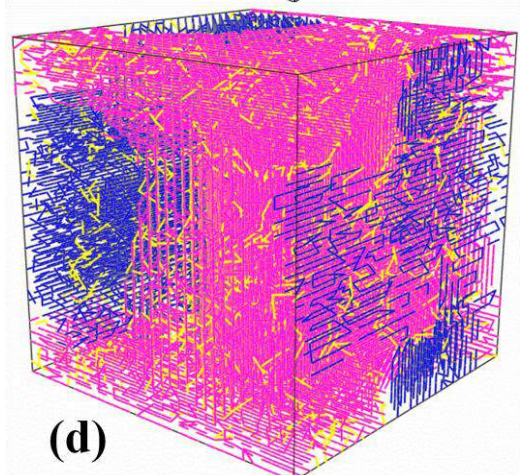

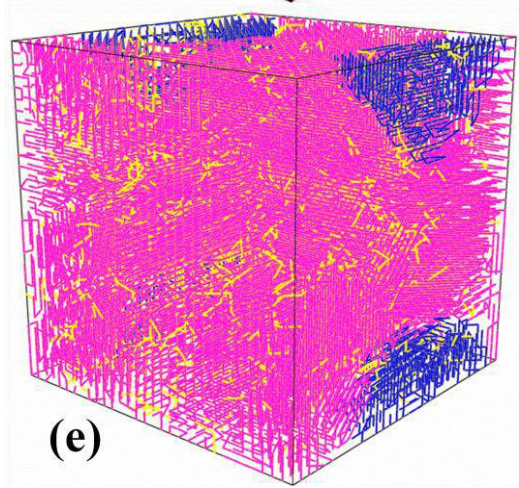

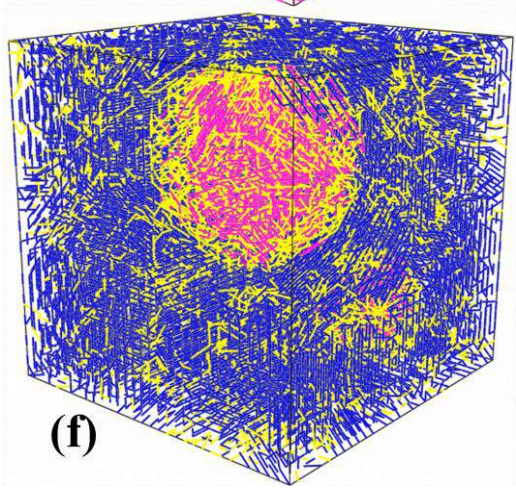

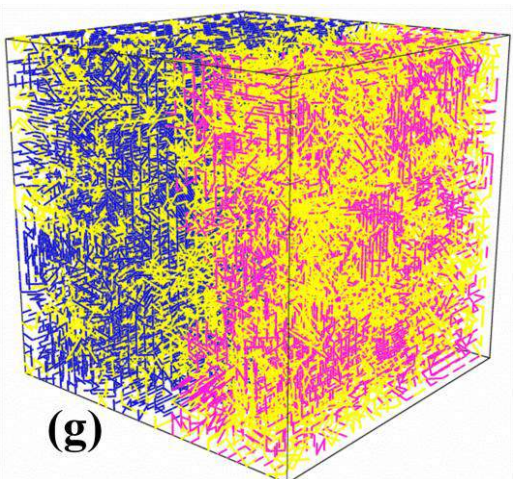

(g)

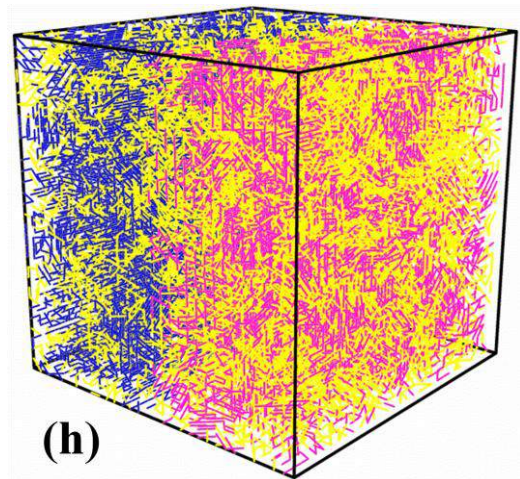

(h)

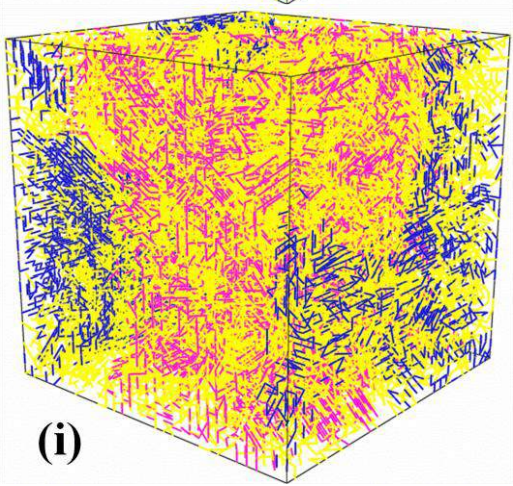

(i)

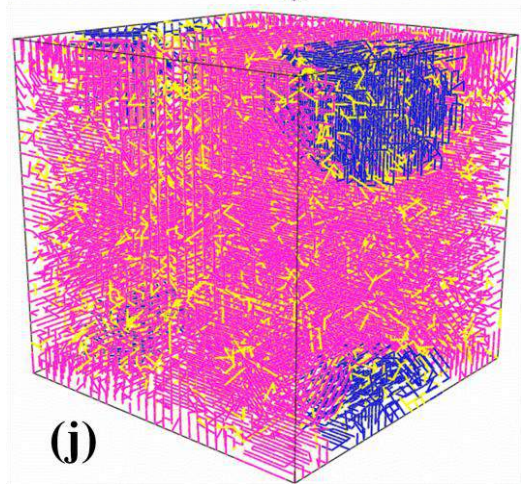

(j)

**Fig. S5 Change in average crystallite size with** $U_p$ **for A- and B-polymers: (a)** $\langle S_A \rangle$ **vs.** $U_p$ **at** $\lambda = 1$, **(b)** $\langle S_A \rangle$ **vs.** $U_p$ **at** $\lambda = 6$, **(c)** $\langle S_B \rangle$ **vs.** $U_p$ **at** $\lambda = 1$, **(d)** $\langle S_B \rangle$ **vs.** $U_p$ **at** $\lambda = 6$.

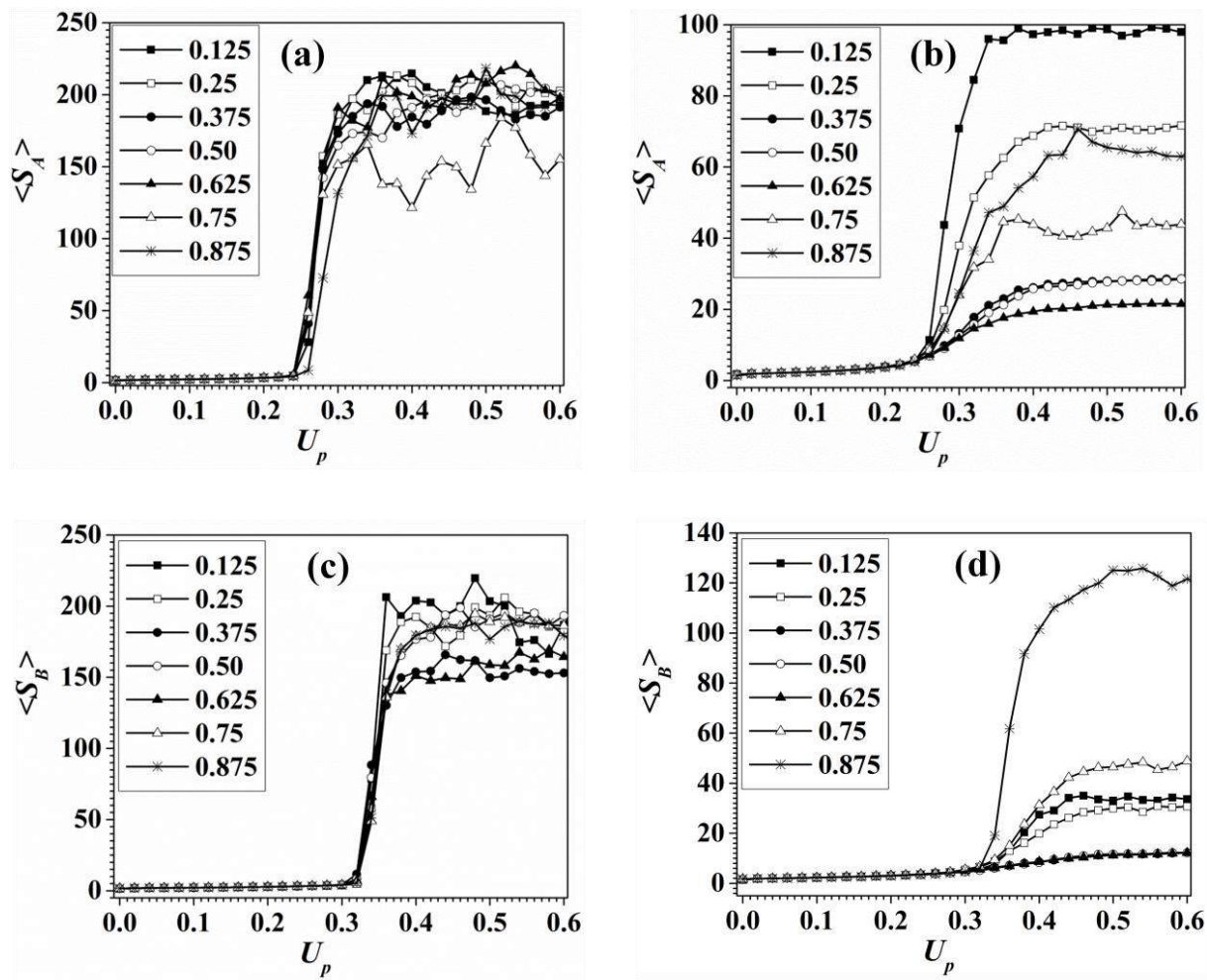

**Fig. S6 Change in average lamellar thickness with** $U_p$ **for A- and B-polymers: (a)** $\langle l_A \rangle$ **vs.** $U_p$ **at** $\lambda = 1$**, (b)** $\langle l_A \rangle$ **vs.** $U_p$ **at** $\lambda = 6$**, (c)** $\langle l_B \rangle$ **vs.** $U_p$ **at** $\lambda = 1$**, (d)** $\langle l_B \rangle$ **vs.** $U_p$ **at** $\lambda = 6$**.**

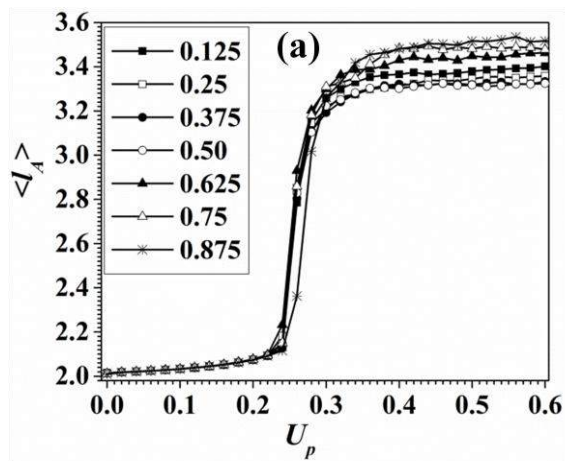
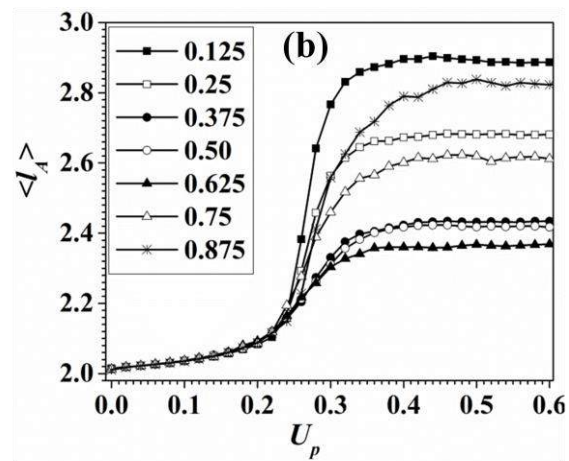
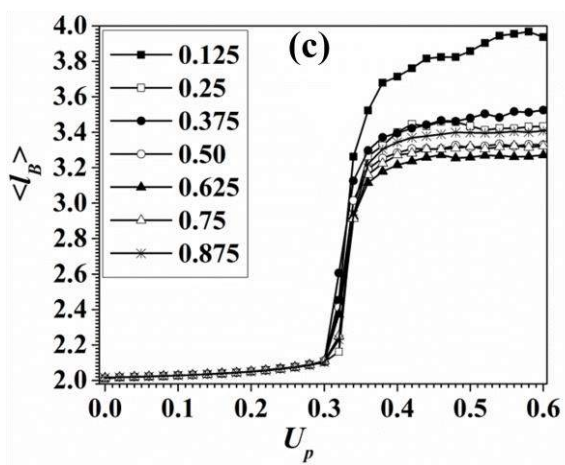
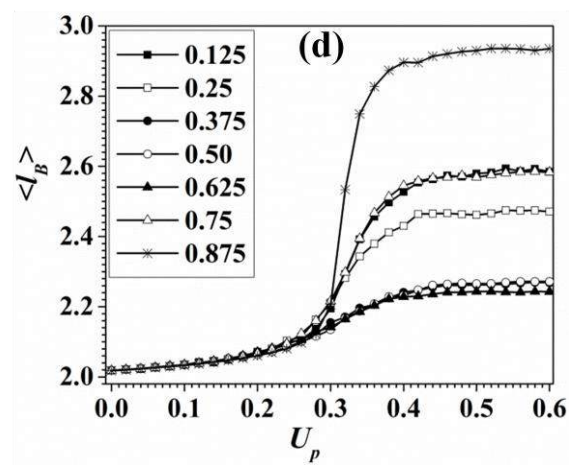

**Table S1 Comparison in saturated crystallinity of A-polymer, $X_A$, B-polymer, $X_A$, with composition, $x_B$, during one -step isothermal crystallization, at $\lambda = 1$ and 6.**

| | Weak segregation, $\lambda = 1$ | | Strong segregation, $\lambda = 6$ | |
|---|---|---|---|---|
| Composition ($x_B$) | $X_A$ | $X_B$ | $X_A$ | $X_B$ |
| 0.125 | 0.683 | 0.632 | 0.175 | 0.053 |
| 0.25 | 0.664 | 0.639 | 0.085 | 0.048 |
| 0.375 | 0.660 | 0.639 | 0.070 | 0.054 |
| 0.5 | 0.653 | 0.641 | 0.072 | 0.066 |
| 0.625 | 0.645 | 0.648 | 0.051 | 0.064 |
| 0.75 | 0.645 | 0.663 | 0.051 | 0.080 |
| 0.875 | 0.641 | 0.690 | 0.067 | 0.167 |

**Table S2 Comparison in average lamellar thickness of A-polymer, $\langle l_A \rangle$, with composition, $x_B$, during one- and two-step isothermal crystallization, at $\lambda = 1$.**

| | Two-step cooling | | One-step cooling |
|---|---|---|---|
| Composition ($x_B$) | $U_p = 0.28$ | $U_p = 0.6$ | $U_p = 0.6$ |
| 0.125 | 3.47 | 3.51 | 2.7 |
| 0.25 | 3.46 | 3.46 | 2.67 |
| 0.375 | 3.39 | 3.44 | 2.68 |
| 0.5 | 3.43 | 3.47 | 2.74 |
| 0.625 | 3.45 | 3.47 | 2.73 |
| 0.75 | 3.53 | 3.63 | 2.81 |
| 0.875 | 3.61 | 3.72 | 2.95 |

**Table S3 Comparison in average lamellar thickness of B-polymer, $\langle l_B \rangle$, with composition, $x_B$, during one- and two-step isothermal crystallization, at $\lambda = 1$.**

| | Two-step cooling | One-step cooling |
|---|---|---|
| Composition ($x_B$) | $U_p = 0.6$ | $U_p = 0.6$ |
| 0.125 | 3.00 | 3.01 |
| 0.25 | 2.83 | 2.85 |
| 0.375 | 2.79 | 2.79 |
| 0.5 | 2.78 | 2.74 |
| 0.625 | 2.77 | 2.74 |
| 0.75 | 2.78 | 2.75 |
| 0.875 | 2.86 | 2.80 |

**Table S4 Comparison in saturated crystallinity of A-polymer, $X_A$, with composition, $x_B$, during one- and two-step isothermal crystallization, at $\lambda = 6$.**

| | Two-step cooling | | One-step cooling |
|---|---|---|---|
| **Composition ($x_B$)** | $U_p = 0.28$ | $U_p = 0.6$ | $U_p = 0.6$ |
| **0.125** | 0.623 | 0.668 | 0.175 |
| **0.25** | 0.397 | 0.473 | 0.085 |
| **0.375** | 0.294 | 0.335 | 0.070 |
| **0.5** | 0.384 | 0.426 | 0.072 |
| **0.625** | 0.358 | 0.422 | 0.051 |
| **0.75** | 0.276 | 0.329 | 0.051 |
| **0.875** | 0.429 | 0.523 | 0.067 |

**Table S5 Comparison in saturated crystallinity of B-polymer, $X_B$, with composition, $x_B$, during one- and two-step isothermal crystallization, at $\lambda = 6$.**

| | Two-step cooling | | One-step cooling |
|---|---|---|---|
| **Composition ($x_B$)** | $U_p = 0.28$ | $U_p = 0.6$ | $U_p = 0.6$ |
| **0.125** | 0.126 | 0.342 | 0.053 |
| **0.25** | 0.144 | 0.243 | 0.048 |
| **0.375** | 0.145 | 0.180 | 0.054 |
| **0.5** | 0.151 | 0.250 | 0.066 |
| **0.625** | 0.143 | 0.259 | 0.064 |
| **0.75** | 0.150 | 0.246 | 0.051 |
| **0.875** | 0.137 | 0.523 | 0.167 |